\documentclass[
  aps,
  prfluids,
  amsmath,
  amsfonts,
  amssymb,
  longbibliography,
  nofootinbib,
]{revtex4-1}
\usepackage[paper=a4paper,textwidth=175mm,textheight=220mm]{geometry}

\usepackage[utf8]{inputenc}
\usepackage{microtype}
\usepackage{xparse}
\usepackage{mathrsfs}

\usepackage{booktabs}
\usepackage[hidelinks]{hyperref}
\usepackage{siunitx}
\usepackage{bm}
\usepackage[normalem]{ulem}

\usepackage{float}
\usepackage{graphicx}
\usepackage[svgnames]{xcolor}
\usepackage{csquotes}

\usepackage{geometry}

\usepackage{color}
\usepackage{gensymb}
\usepackage{bbm}

\usepackage{array,multirow}

\newcolumntype{R}[1]{>{\raggedleft\arraybackslash }b{#1}}
\newcolumntype{L}[1]{>{\raggedright\arraybackslash }b{#1}}
\newcolumntype{C}[1]{>{\centering\arraybackslash }b{#1}}

\newcommand\ve[1]{\boldsymbol{#1}}
\newcommand{\ma}[1]{\ensuremath{\mathbb{#1}}}
\newcommand{\ku}{{\rm Ku}}
\newcommand{\st}{{\rm St}}
\newcommand{\sv}{{\rm Sv}}
\newcommand{\gr}{{ g}}

\newcommand{\fr}{{\rm Fr}}
\newcommand{\rep}{{\rm Re}_{\rm p}}
\newcommand\Ao{A_\perp}
\newcommand\Ap{A_\parallel}
\newcommand\Co{C_\perp}
\newcommand\Io{I_\perp}
\newcommand\Cp{C_\parallel}
\newcommand\Ip{I_\parallel}
\newcommand\dphi{\delta\varphi}
\newcommand{\nnhat}{\ensuremath{\hat{\ve n}}}
\newcommand{\pphat}{\ensuremath{\hat{\ve p}}}
\newcommand{\sshat}{\ensuremath{\hat{\ve s}}}
\newcommand{\gghat}{\ensuremath{\hat{\ve g}}}
\newcommand{\dissip}{\varepsilon}

\begin{document}

\title{Statistical orientation and distribution of columnar ice crystals in turbulent flows}
\author{Alain Pumir$^{1,2}$}
\author{Muhammad Zubair Sheikh$^3$}
\author{Kristian Gustavsson$^4$}
\author{Emmanuel L\'ev\^eque$^5$}
\author{Bernhard Mehlig$^4$}
\author{Aurore Naso$^5$}

\affiliation{
$^1$CNRS, Ecole Normale Superieure de Lyon, Laboratoire de Physique, UMR5672, F-69007, Lyon, France\\
$^2$Max Planck Institute for Dynamics and Self-Organization, G\"ottingen, D-37077, Germany\\
$^3$Department of Mechanical Engineering, University of Engineering and Technology Lahore, 54890, Lahore, Pakistan\\
$^4$Department of Physics, Gothenburg University, 41296 Gothenburg, Sweden \\
$^5$CNRS, Ecole Centrale de Lyon, INSA Lyon, Universite Claude Bernard Lyon 1, Laboratoire de M\'ecanique des Fluides et d'Acoustique, UMR5509, F-69134, Ecully, France
}

\begin{abstract}
We study the motion of columnar ice crystals that form in clouds over a range
of low temperature. Our focus here is on elongated ice crystals,
which are
smaller than the size of the smallest eddies in the flow, with a moderate
aspect ratio comprised between $3$ and $5$.  We determine
turbulent solutions of the Navier-Stokes equations over a
range of turbulent kinetic energy dissipation characteristic of clouds
($4.41\;{\rm cm}^2/{\rm s}^3 \le \dissip \le 1120\;{\rm cm}^2/{\rm s}^3$) by
using direct numerical simulations, and we follow the motion of crystals using
simplified but realistic models for the motion of non-spherical, elongated
particles.
The influence of the fluid inertia leads to a preferential alignment of the
crystals perpendicular to the direction of gravity, the alignment effect
being opposed by the turbulent fluctuations.
Along with the strong alignment of the crystal axis
perpendicular to gravity, we observe
only a weak alignment with the vorticity, much weaker than in the 
absence of gravity.
The settling velocity depends
only weakly on the orientation of the crystals, but is strongly enhanced when
$\dissip$ increases, an effect that we attribute to preferential
concentration in the flow.  As the inertia of the columnar ice crystals 
considered here
is significant, we observe a strong spatial clustering. Finally, we discuss the
relevance of the effects identified here on the collision 
frequency between
ice crystals in cloud conditions.

\end{abstract}

\maketitle

\section{Introduction} \label{sec:introduction}

In cold clouds, pristine crystals of various shapes may form, depending on the external
conditions~\cite{Pru78}. In particular, columnar
ice crystals prevail either in very cold
conditions ($T \le -22 \;^\circ {\rm C} $) or in warmer conditions
($-8 \;^\circ {\rm C} \le T \le -3 \;^\circ {\rm C}$)~\cite{Pru78,LambVerlinde}.
These crystals settle through air, and in quiescent conditions align with their broad
sides facing downwards. This may lead to very spectacular halos~\cite{Moilanen:2022}.
Turbulence tends to randomize particle orientations, 
therefore weakening
the alignment effect. Even in the presence of moderate turbulent conditions, 
however, the
alignment of crystals contributes to the reflection of radiations,
therefore potentially affecting the albedo of the earth, 
with significant
climatic implications~\cite{Noel2010,Saito19}. The growth of ice crystals, through
collisions with other ice crystals or with small droplets present in the cloud,
is also essential to understand the formation of precipitating hydrometeors~\cite{Pru78}.
This work is devoted to understanding the alignment of columnar ice crystals as they settle
through turbulent clouds.

Describing the motion of particles in a turbulent environment is an essential aspect of
cloud microphysics, and has therefore received considerable
attention~\cite{Pru78}, among others from a numerical point of view 
as recently reviewed e.g. in~\cite{Wang_2021}. In our study, we restrict
ourselves to small columnar ice crystals, and study their motion as they settle in
a turbulent cloud. The present study is complementary to our recent work, devoted
mostly to the motion of small spheroids with disk-like
shape~\cite{Siew14a,Gus17,Jucha2018,Gus19,She20}.
To study the motion of non-spherical particles, one must determine the force and torque
due to the interaction with the fluid. The fact that the 
particles considered
in our studies are relatively small, and are therefore moving in the fluid at a small
particle Reynolds number $\rep$, may suggest that it is
sufficient to consider the force and torque acting on the particle in the Stokes regime
($\rep \to 0$). However, the corresponding approximation~\cite{Siew14a,Gus17}
leads to incorrect predictions for the alignment of settling particles.
The importance of fluid inertia to capture qualitatively the orientation of small anisotropic
particles settling in a flow, even when the particle Reynolds number is of order $\approx 1$,
is an essential lesson of recent studies~\cite{Lop17,Kramel,Men17,Gus19,She20,Roy2023}.
The expression for the torque can be established by perturbation 
for $0 <  \rep \ll 1$~\cite{Cox65,Kha89,Dab15}. We 
use here the theoretical expression for the torque
derived analytically by~\cite{Dab15}, which can be
extended to values of $\rep$ up to $\sim 20$ by introducing
empirically determined dependencies of the parameters 
on $\rep$~\cite{Froh20,Ouch20}
to parametrize the effect of finite $\rep$.
The expression for the torque acting on the particles
has been validated experimentally~\cite{Roy:2019,Cabrera22}, and
the model successfully predicts the
dynamics of
spheroidal particles settling in still fluid~\cite{Bhowmick24}.
Here, we utilize the model, incorporating the physical elements briefly introduced above,
to study the motion of spheroidal particles in a turbulent flow with the help
of direct numerical simulations (DNS) of the Navier-Stokes equations. In technical
terms, we restrict ourselves to the one-way-coupling approach, consisting in
neglecting the feedback of particles on the flow.

Several numerical studies have considered the motion of rod-like particles in a turbulent
flow. Remarkably, in the absence of any inertia and of settling, rods tend to align with the
direction of vorticity~\cite{Pum11}. In the presence of gravity, fluid inertia plays
a crucial role in aligning particles with respect to the direction of
settling~\cite{Lop17,She20,Ana20}. Turbulence, however, tends to randomize the distribution
of orientation, resulting in a competition between the fluid inertia and the turbulent
fluctuations~\cite{Kle95,Kramel,Men17,Gus19,Gus21}. In particular, the predictions of~\cite{Gus21} for
settling disk-like particles agree very well with DNS.

In the case of disk-shaped particles, the settling velocity significantly depends on the angle
between the axis of the particle and gravity. This property results from the dependence of
the drag on the orientation of the particle with respect to the flow. It affects
the collision rate between oblate particles, particularly at moderate turbulence
intensity~\cite{She22}.

The present study is motivated by the motion of columnar ice crystals in a turbulent flow.
We are restricting ourselves to a range of parameters 
relevant to cloud dynamics~\cite{Pru78}, and
we study the elementary properties of motion of columnar crystals
settling in a turbulent flow.
Specifically, detailed studies of the shape of columnar ice crystals have suggested simple empirical
relation between the length of crystals, $L$, and their diameter, $d$, which
depend on the precise type of the crystal. An example is given by:
\begin{align}
d/d_0 =  (L/d_0)^{0.437} ~~~ {\rm with} ~~~ d_0 = 2.63 \times 10^{-3} \, {\rm cm} \,  ,
\end{align}
for long solid columns (crystals N1e~\cite{Jayaweera:1974}, see also 
Table 2.2b in~\cite{Pru78}).
The corresponding aspect ratio, defined as $\beta = \frac{L}{d}$, is given by
$\beta \approx (L/d_0)^{0.563}$.
We approximate the shape of a crystal as a
spheroid, and we consider three sets of parameters, with $\beta = 3$, $4$ and $5$, corresponding respectively to
$L = 200\;\mu {\rm m}$, $300\;\mu {\rm m}$ and $400\;\mu {\rm m}$.
We also vary the turbulence intensity, via the energy dissipation rate $\dissip$.
Here, we take $\dissip$ in a range typical for cloud turbulence:
$4.43\;{\rm cm}^2/{\rm s}^3 \le \dissip  \le 1120\;{\rm cm}^2/{\rm s}^3$.  \\
An important property of these particles is that their dynamics involves a
relaxation time.
In the range of parameters covered in our study, the particle inertia
is significantly larger than in~\cite{Gus21,She22}, and is large enough to induce
pronounced spatial inhomogeneities of the
particle distribution in the flow.
Accordingly, we extend the analysis in~\cite{Gus21,She22} by examining
the alignment of rod-like 
particles not only with
gravity,
but also with vorticity and with the strain eigendirections. Furthermore,
we investigate the particle distribution in the flow.

The work is presented as follows. In section~\ref{sec:num_setup}, 
we summarize the
characteristics of our DNS flows, and introduce the model equations used to describe
the dynamics of the particles. Section~\ref{sec:results} contains our main results. We
first discuss, see Subsection~\ref{subsec:res_orientation}, the orientation of particles with
respect to the direction of settling. Subsection~\ref{subsec:sett_vel} is devoted to the
settling velocity of the particles and in particular on its dependence on the orientation
of the particles with respect to the direction of settling. The issue of alignment of
particles with the characteristic directions of the velocity gradient tensor is discussed
in Subsection~\ref{subsec:alignt_A}. The finite inertia of the particles
investigated in this work results in significant spatial heterogeneities of the distribution
of particles, which we study in Subsection~\ref{subsec:distribution}. Finally,
we summarize our results and present our conclusions in Section~\ref{sec:conclusion}.

\begin{table}[b!]
\begin{center}
\begin{tabular}{|C{2.5cm}|C{2cm}|C{2cm}|C{2cm}|C{2cm}|}
\hline
{\bf Flow} & F1 & F2 & F3 & F4\\
\hline
$\dissip$ (${\rm cm}^2 \, {\rm s}^{-3}$) & $4.43$ & $17.8$ & $71.0$ & $1120$\\
\hline
$R_\lambda$ & $56$ & $75$ & $95$ & $150$\\
\hline
$\eta$ (${\rm cm}$) & $0.196$ & $0.139$ & $9.82\;10^{-2}$ & $4.93\;10^{-2}$\\
\hline
$\fr$ & $4.73 \; 10^{-3}$ & $1.34 \; 10^{-2}$ & $3.79\;10^{-2}$ & $3.00\;10^{-1}$\\
\hline
$N$ & $128$ & $128$ & $256$ & $512$\\
\hline
$k_{max}\eta$ & $3.1$ & $2.2$ & $3.1$ & $3.1$\\
\hline
\end{tabular}
\caption{Properties of the four turbulent flows used in this study: energy dissipation/forcing rate $\dissip$, Reynolds 
number based on the Taylor microscale, $R_\lambda$, Kolmogorov scale $\eta$,
Froude 
 number defined by Eq.~\eqref{eq:St_Sv}, number of collocation points in one direction $N$. $k_{max}$ is the largest wavenumber resolved. In all these flows, the values of the fluid kinematic viscosity and density are respectively equal to $\nu=0.1875\;{\rm cm}^2.{\rm s}^{-1}$ and $\rho_f=9.022\;10^{-4}\;{\rm g}.{\rm cm}^{-3}$.}
\label{tab:flows}
\end{center}
\end{table}

\section{Numerical setup} \label{sec:num_setup}

The choice of parameters in the 
present study is motivated by cloud microphysics
of small columnar ice crystals, as introduced in 
Section~\ref{sec:introduction}.
For this reason, we vary separately particle size and
energy dissipation in the turbulent flow. As a consequence,
the Stokes number, describing particle inertia and defined below (see Eq.~\eqref{eq:St_Sv}), is not varied 
independently of the turbulence intensity 
(the Reynolds number). Consequently,
particle inertia increases with the
turbulence intensity at a fixed particle geometry. In this sense, 
our study differs in an essential way from other studies
that varied these two parameters independently (see, {\it e.g.}, ~\cite{Bec:24}).

\subsection{Turbulent flow}

We consider here statistically homogeneous and isotropic turbulent flows.
This choice is a modeling assumption, which can
be rationalized as follows. The particles considered in this 
work are all significantly smaller than the smallest size of the
eddies (the Kolmogorov scale, $\eta$, defined below), so the properties of 
alignment we are considering result essentially
from the interaction between small-scale turbulence and other effects,
such as inertia or gravity. Furthermore, an essential tenet of turbulence
theory is that the small-scale properties of the flow are universal, 
independently of the specificities of the flow at the largest scales of the 
flow~\cite{Fri97,Buaria:25}, and possibly of the temperature and moisture 
stratification.
The present work uses direct numerical simulation of the incompressible Navier-Stokes equations:
\begin{eqnarray}
\frac{\partial {\bf u}}{\partial t}+({\bf u}\cdot{\bf \nabla}){\bf u}=-\frac{1}{\rho_f}{\bf \nabla}p+\nu{\bf \nabla}^2{\bf u}+{\bf F}, \label{eq:NS_1}\\
{\bf \nabla}\cdot{\bf u}=0, \label{eq:NS_2}
\end{eqnarray}
where ${\bf u}({\bf x},t)$ and $p({\bf x},t)$ are respectively the velocity and the pressure fields, $\rho_f$ is the fluid density, and $\nu$ is its kinematic viscosity. The term ${\bf F}({\bf x},t)$ is a force which continuously injects a fixed amount of kinetic energy per unit mass, $\dissip$,
into the flow \cite{Lamorgese04}, thereby maintaining a statistically steady state.

The Navier-Stokes equations (\ref{eq:NS_1},\ref{eq:NS_2}) were integrated in a triply periodic domain of size $L_{\rm box}^3$, where $L_{\rm box}=2\pi$ in dimensionless units. The solver uses a standard Fourier pseudospectral method, including the 2/3 dealiasing rule to suppress discretization errors stemming from the nonlinear term.  \\
In physical terms, we scale the spatial and temporal units, so the box size in physical units is $8 \pi \; {\rm cm}$, and the code time unit is $1\; {\rm s}$. The simulations were performed for a fixed value of $\nu$ and four values of the energy dissipation rate $\dissip$, on $N^3$ collocation points, giving rise to four turbulent flows respectively denoted as F1, F2, F3 and F4. The values of these parameters have been chosen to reflect the microphysics of clouds at $-5\;\degree C$ in a box of size $8\pi\approx 25\;{\rm cm}$ in each direction. The flow characteristics can be found in Tab. \ref{tab:flows}. The largest resolved wavenumber, $k_{max}$, satisfies the stringent criterion $k_{max}\eta >3$ for flows F1, F3 and F4, where $\eta=(\nu^3/\dissip)^{1/4}$ is the Kolmogorov scale. This very high resolution of the smallest scales is due to the fact that the particles motion depends on the flow velocity gradients, which must be interpolated at the particles location. The value of $k_{max}\eta$ is slightly smaller for flow F2, but we have checked that the results obtained in this flow remain valid with a resolution of $256^3$. Further details about the code can be found in \cite{Jucha2018,Naso2018,She20,She22,She24}.
As a caveat, we notice that, although the values of $\dissip$ are
consistent with those observed in clouds, the Reynolds numbers in the 
atmosphere are much larger, due to the sizes involved (several hundred
of meters),
vastly superior to the small size of our box ($L \approx 25 cm$).

\subsection{Particle dynamics}

The particles considered in the present study are prolate ellipsoids of revolution of density $\rho_p$. The geometry of these elongated spheroids is characterized by their semi-major axis $c=L/2$ and their aspect ratio $\beta>1$, associated to a semi-minor axis $a=d/2=c/\beta$. As already mentioned, the particle properties have been chosen to represent at best small ice crystals in clouds. We chose here parameters corresponding to a temperature of $-5\;\degree C$ \cite{Pru78,Ramelli24}; but focusing on the colder regime where columnar crystals
form, for $T < - 22 \;^\circ C$, would not change the conclusions of our work.
Three particle types, whose characteristics are provided in Tab. \ref{tab:particles}, were therefore considered. The major axis $2c$ is always smaller than $\eta$ (see Tab. \ref{tab:flows} and \ref{tab:particles}). We use the one-way coupling approximation, according to which the particle feedback on the flow and the particle-particle interactions are neglected.

\begin{table}[b!]
\begin{center}
\begin{tabular}{|C{2cm}|C{2cm}|C{2cm}|C{2cm}|C{2cm}|C{3cm}|C{2cm}|}
\hline
{\bf Particle} & $c$ ($\mu {\rm m}$) & $\beta$ & $a$ ($\mu {\rm m}$) & $\rho_p$ (${\rm g} \, {\rm cm}^{-3}$) & $m$ ($\times 10^7\;{\rm g}$)\\
\hline
P1 & $100$ & $3$ & $33.3$ & $0.8673$ & $4.03$\\
\hline
P2 & $150$ & $4$ & $37.5$ & $0.8624$ & $7.62$\\
\hline
P3 & $200$ & $5$ & $40$ & $0.8589$ & $11.5$
\\
\hline
\end{tabular}
\caption{Properties of the three particle types used in this study: semimajor axis $c$, aspect ratio $\beta$, semiminor axis $a=c/\beta$, density $\rho_p$, and mass $m=\frac{4}{3}\pi\rho_p a^2c$.}
\label{tab:particles}
\end{center}
\end{table}

\begin{figure}[b!]
    \centering
            \includegraphics[trim=3cm 9cm 4cm 9cm,clip,height=5.9cm,keepaspectratio]{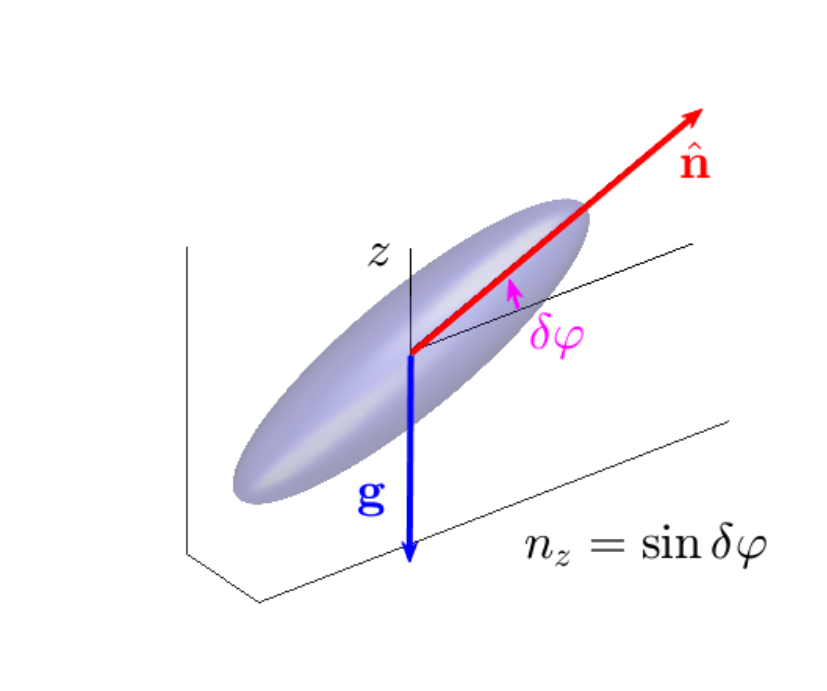}
    \caption{Characterization of the particle orientation: $\hat{\bf n}$ is a unit vector collinear with the axis of symmetry of the particle, $n_z$ is the projection of $\hat{\bf n}$ on the gravitational acceleration ${\bf g}$, and $\delta\varphi$ is the angle between $\hat{\bf n}$ and the horizontal direction.}
	\label{fig:n}
\end{figure}

The motion of a particle is described by Newton's equations for the position of its center of mass $\mathbf{x}$ and its velocity $\mathbf{v}$, as well as for its orientation $\hat{\mathbf{n}}$, defined as a unit vector parallel to the axis of symmetry of the particle (Fig.~\ref{fig:n}), and its angular velocity $\bm{\omega}$:
\begin{equation}
\frac{d \mathbf{x}}{dt} = \mathbf{v} ~, ~~~ m \frac{d \mathbf{v}}{dt} = \mathbf{f}_h + m \mathbf{g},
\label{eq:Newt_transl}
\end{equation}
where $\mathbf{f}_h$ is the hydrodynamic force acting on the object, and
\begin{equation}
\frac{d \hat{\mathbf{n}}}{dt} = \bm{\omega} \times \hat{\mathbf{n}}, \quad m \frac{d}{dt} \left[ {\mathbb I} ( \hat{\mathbf{n}} ) \bm{\omega} \right] = \bm{\tau}_h,
\label{eq:Newt_orient}
\end{equation}
where $\bm{\tau}_h$ is the hydrodynamic torque acting on it.
In Eqs.~\eqref{eq:Newt_transl} and \eqref{eq:Newt_orient}, ${\bf g}$ is the gravitational acceleration ($\gr=|{\bf g}|=981\;{\rm cm}/{\rm s}^{2}$), $m$ denotes the mass of the particle and ${\mathbb I} (\hat{\mathbf{n}})$ its moment of inertia tensor given in Appendix \ref{app:appendix}. In Eq. (\ref{eq:Newt_transl}), the buoyancy force has been neglected with respect to the particle weight, since $\rho_p \gg \rho_f$. We notice that the 
particles considered in this work are highly symmetric, so the geometric 
center coincides with the centers of resistance and buoyancy.  \\

The difficulty lies in the modelling of the hydrodynamic force and torque, $\mathbf{f}_h$ and $\bm{\tau}_h$. Following previous works \cite{Gus19,She20,Gus21,She22,She24}, we adopt here a simplified model \cite{Kle95,Kramel,Lop17} in which small inertial corrections due to convective fluid inertia are added to the standard expressions for $\mathbf{f}_h$ and $\bm{\tau}_h$ in the creeping-flow limit. In the regimes investigated in the present work, the particle Reynolds number $\rep =2c|{\bf v}-{\bf u}|/\nu$, based on the relative velocity of the particle in the flow (${\bf v}$ and ${\bf u}$ respectively denote the particle velocity and the fluid velocity at the same position) and its major axis, is of order $1-10$. It was recently shown that the force and torque can be parametrized, for the highly symmetric objects considered in our study, by introducing empirically determined correction factors in the perturbative equations of motion. We therefore use here the following expressions:
\begin{equation}
\mathbf{f}_h=\mathbf{f}_h^{(0)}+C_f\mathbf{f}_h^{(1)};\quad \bm{\tau}_h=\bm{\tau}_h^{(0)}+C_\tau\bm{\tau}_h^{(1)}, \label{eq:torque_sum}
\end{equation}
with correction factors $C_f=0.6$ and $C_\tau=0.5$ \cite{Bhowmick24}.\\

In the creeping-flow limit, the hydrodynamic force is simply Stokes' force:
\begin{equation}
\mathbf{f}_h^{(0)} = 6\pi a\mu {\mathbb A} \big({\bf u}-{\bf v}), \label{eq:f0}
\end{equation}
where ${\mathbb A}$ is a resistance tensor relating $\mathbf{f}_h^{(0)}$ and the slip velocity ${\bf v}-{\bf u}$ \cite{Kim:2005}, and the hydrodynamic torque is Jeffery's torque \cite{Jef22}:
\begin{equation}
\bm{\tau}_h^{(0)} =  6\pi a\mu [{\mathbb C}(\boldsymbol{\Omega}-\boldsymbol{\omega})+{\mathbb H}\cdot{\mathbb S}], \label{eq:tau0}
\end{equation}
where $\boldsymbol{\Omega}=\tfrac{1}{2} \boldsymbol{\nabla} \wedge {\bf u}$ is half the fluid vorticity at the particle position, $\boldsymbol{\omega}-\boldsymbol{\Omega}$ is the angular slip velocity, and the tensors ${\mathbb C}$ and ${\mathbb H}$ determine the coupling of the hydrodynamic torque with vorticity $\boldsymbol{\Omega}$ and strain ${\mathbb S}$. The three tensors ${\mathbb A}$, ${\mathbb C}$ and ${\mathbb H}$ depend on the orientation $\hat{\mathbf{n}}$. Their expressions are provided in Appendix \ref{app:appendix}.

The steady convective-inertia corrections to the force and torque in a quiescent fluid at leading order in the particle Reynolds number $\rep$ are \citep{Brenner61,Cox65,Kha89,Dab15}:
\begin{subequations}
\label{eq:inertial_corrections}
\begin{align}
 \label{eq:drag_correction}
&\hspace*{-2mm}\mathbf{f}_h^{(1)} \!=\! -(6\pi a\mu )\frac{3}{16}\frac{aW}{\nu} \big[3{\mathbb A}\!-\!
\mathbbm{1}
(\hat{\bf W}\!\cdot {\mathbb A} \hat {\bf W})\big]{\mathbb A} {\bf W}, \\
&\hspace*{-2mm}\bm{\tau}_h^{(1)}= F(\beta){\mu}   \,\frac{a^3W^2}{\nu}\,
 (\hat{\bf n}\cdot {\hat{\bf W}})(\hat{\bf n}\wedge {\hat{\bf W}})\,. \label{eq:tau1}
\end{align}
\end{subequations}
Here $W = |\bf W|$ is the modulus of the slip velocity ${\bf W}={\bf v}-{\bf u}$ and $\hat{\bf W} = {\bf W}/W$ is its direction, $\mathbbm{1}$ is the unit matrix, and $F(\beta)$ is a shape factor computed for spheroids of arbitrary aspect ratio by \cite{Dab15} and whose expression is provided in Appendix \ref{app:appendix}.\\

The fluid-inertia correction to the torque also includes in principle a contribution due to the fluid shear $s$. The expression of this contribution in a turbulent flow is still unknown, but it is expected to scale at first order as $Re_s^{1/2}$, where $Re_s=a^2s/\nu$ is the particle Reynolds number based on the local shear rate \cite{Candelier19}. In a turbulent flow, $Re_s\sim (a/\eta)^2$ since $s\sim 1/\tau_\eta$. For particles smaller than the Kolmogorov scale, the inertial correction due to shear
is therefore negligible compared to Jeffery’s torque.\\

The dynamics of a particle settling in a turbulent flow depends on two dimensionless parameters \cite{Dev12}: the Stokes number $\st$, defined as the ratio between the particle response time $\tau_p$ and the Kolmogorov time scale $\tau_\eta=(\nu/\dissip)^{1/2}$, and the settling number $\sv=\gr \tau_p/u_\eta$, defined as the ratio between the particle settling velocity in a quiescent flow and the Kolmogorov velocity $u_\eta=(\nu\dissip)^{1/4}$. 
Consistent e.g. with~\cite{Gus21}, we use the expression $\tau_{\rm p} = 2 a^2 \beta/(9 \nu) (\rho_p/\rho_f)$. The expression used in~\cite{She20}, 
$\tau_{\rm p}' = a^2\log\beta/(3\nu) (\rho_p/\rho_f )$, more appropriate to study particles with a value of $\beta \gg 1$, differs by no more than a factor of $2$
in the range of values of $\beta$ considered here.
As a consequence, the values of $\st$ and $\sv$ used in this work are:
\begin{equation}
\st=\frac{2\beta}{9} \left( \frac{a}{\eta} \right) ^2 \frac{\rho_p}{\rho_f},\quad \sv=\frac{2}{9} \frac{\gr a^2\beta}{(\nu^5\dissip)^{1/4}}\frac{\rho_p}{\rho_f}  = \frac{\st}{\fr} ~ , ~~~ {\rm where} ~~ \fr = \frac{\dissip^{3/4}/\nu^{1/4}}{\gr } \, .
\label{eq:St_Sv}
\end{equation}
In Eq.~\eqref{eq:St_Sv}, we have re-expressed $\sv$ as the ratio between $\st$ and the Froude number, defined as $\fr = \langle \mathbf{a}^2 \rangle^{1/2}/\gr $, where $\langle \mathbf{a}^2 \rangle^{1/2}$ is a measure of the acceleration in the fluid. Standard estimates lead to
$\langle \mathbf{a}^2 \rangle^{1/2} = \dissip^{3/4}/\nu^{1/4}$~\cite{Heisenberg48,Yaglom49}. The values of the Froude numbers for the four flows studied here are indicated in Table~\ref{tab:flows}.

\begin{table}[t!]
\begin{center}
\begin{tabular}{|C{2cm}|C{2cm}|C{2cm}||C{2cm}|C{2cm}||C{2cm}|C{2cm}|}
\hline
Flow & Particle & $\gr$ (${\rm cm}.{\rm s}^{-2}$) & $\st$ & $\sv$ & $\langle (\delta\varphi)^2 \rangle ^{1/2}\;(^\circ)$ & $\langle \rep \rangle$\\
\hline
F1 & P1 & $981$ & $0.18$ & $39$ & $1.77$ & $2.0$\\
\hline
F1 & P1 & $0$ & $0.18$ & $0$ & - & $0.020$\\
\hline
F1 & P2 & $981$ & $0.31$ & $65$ & $2.15$ & $4.0$\\
\hline
F1 & P2 & $0$ & $0.31$ & $0$ & - & $0.042$\\
\hline
F1 & P3 & $981$ & $0.44$ & $93$ & $2.31$ & $6.1$\\
\hline
F1 & P3 & $0$ & $0.44$ & $0$ & - & $0.068$\\
\hline
F2 & P1 & $981$ & $0.37$ & $28$ & $3.47$ & $2.0$\\
\hline
F2 & P1 & $0$ & $0.37$ & $0$ & - & $0.057$\\
\hline
F2 & P2 & $981$ & $0.63$ & $47$ & $3.92$ & $4.0$\\
\hline
F2 & P2 & $0$ & $0.63$ & $0$ & - & $0.12$\\
\hline
F2 & P3 & $981$ & $0.89$ & $66$ & $4.36$ & $6.1$\\
\hline
F2 & P3 & $0$ & $0.89$ & $0$ & - & $0.19$\\
\hline
F3 & P1 & $981$ & $0.74$ & $19$ & $6.40$ & $2.1$\\
\hline
F3 & P1 & $0$ & $0.74$ & $0$ & - & $0.15$\\
\hline
F3 & P2 & $981$ & $1.3$ & $33$ & $7.26$ & $4.0$\\
\hline
F3 & P2 & $0$ & $1.3$ & $0$ & - & $0.30$\\
\hline
F3 & P3 & $981$ & $1.8$ & $46$ & $7.71$ & $6.1$\\
\hline
F3 & P3 & $0$ & $1.8$ & $0$ & - & $0.47$\\
\hline
F4 & P1 & $981$ & $2.9$ & $9.8$ & $21.2$ & $2.3$\\
\hline
F4 & P1 & $0$ & $2.9$ & $0$ & - & $0.88$\\
\hline
F4 & P2 & $981$ & $4.9$ & $16$ & $21.9$ & $4.4$\\
\hline
F4 & P2 & $0$ & $4.9$ & $0$ & - & $1.6$\\
\hline
F4 & P3 & $981$ & $7.0$ & $23$ & $22.5$ & $6.7$\\
\hline
F4 & P3 & $0$ & $7.0$ & $0$ & - & $2.4$\\
\hline
\end{tabular}
\caption{List of the simulations performed. $\gr$ denotes the gravitational acceleration, $\st$ is the particle Stokes number, $\sv$ the settling parameter, both defined in Eq. (\ref{eq:St_Sv}), $\langle (\delta\varphi)^2 \rangle ^{1/2}$ is the rms of the angle between the particle axis of symmetry and the horizontal direction (only relevant in the presence of gravity), and $\langle \rep \rangle$ is the mean particle Reynolds number.}
\label{tab:runs}
\end{center}
\end{table}

\subsection{Description of the runs}

A series of simulations, corresponding to the four turbulent flows described in Tab. \ref{tab:flows} and the three particle types characterized in Tab. \ref{tab:particles}, were carried out. In each case, simulations were performed both in the presence and in the absence of gravity. In practice, we simulated a million trajectories for each type of particle and flow, over a time of order $50\;T_e$ for flow F1, $30\;T_e$ for F3, and $20\;T_e$ for F4, where $T_e$ is the large eddy turnover time, defined as the ratio between the integral length scale and the velocity fluctuations.
The particle statistics were calculated at least $50\tau_\eta$ after their injection into the flow, to make sure that they had reached a statistically steady state. The values of $\st$, $\sv$ and of the mean particle Reynolds number $\langle \rep \rangle$ for the resulting 24 runs are given in Tab. \ref{tab:runs}. It is worth mentioning that $\st$ and $\sv$ can be calculated a priori, whereas the values of $\langle \rep \rangle$ were extracted from our simulations.\\

As the flow is periodic in all three dimensions, we chose to reinject
particles leaving the computational domain through one of the faces
directly on the opposite face.
As a consequence, the particle statistics may be artificially influenced by periodicity, particularly in the vertical direction when gravity is
strong~\cite{Ireland16}. Quantitatively, periodicity is expected to
lead to nonphysical effects when the settling time of a particle though
the domain is smaller than the decorrelation time of turbulence \cite{Woittiez09}, {\it i.e.} when $L_{\rm box}/U_{\rm sett} \lesssim T_e$, where $U_{\rm sett}$ is the mean settling velocity of the particle. In our simulations, the characteristic time for
particles to settle through the box was found to be larger than
$2.2\;T_e$ for flow F3, and larger than $5.4\;T_e$ for F4. However, for particles P2 and P3, $(L_{\rm box}/U_{\rm sett})/T_e$ was $\sim 1.4$ in flow F2 and $\sim 0.8$ in F1.

To rule out any spurious effects of periodicity, we compared the values of $U_{\rm sett}$ and the statistics of alignment,
by studying the fluctuations of $n_z$, the projection of $\hat{\bf n}$ on the vertical axis, in computational domains of vertical size $L_{\rm box}$ and $2 L_{\rm box}$.
For particles P2 and P3 in flows F1 and F2, the statistics obtained in both configurations differed by no more than $1\;\%$ for $U_{\rm sett}$, and than $2\;\%$ for the fluctuations of $n_z$.
This allows us to conclude that the particle statistics are not affected by periodicity in the cubic domain. The results presented in such a domain are presented in the next section.

\section{Results} \label{sec:results}

We present and discuss in this section the results of the numerical simulations described in Sec. \ref{sec:num_setup}.

\begin{figure}[t!]
    \centering
            \includegraphics[trim=3cm 9cm 4cm 9cm,clip,height=5.9cm,keepaspectratio]{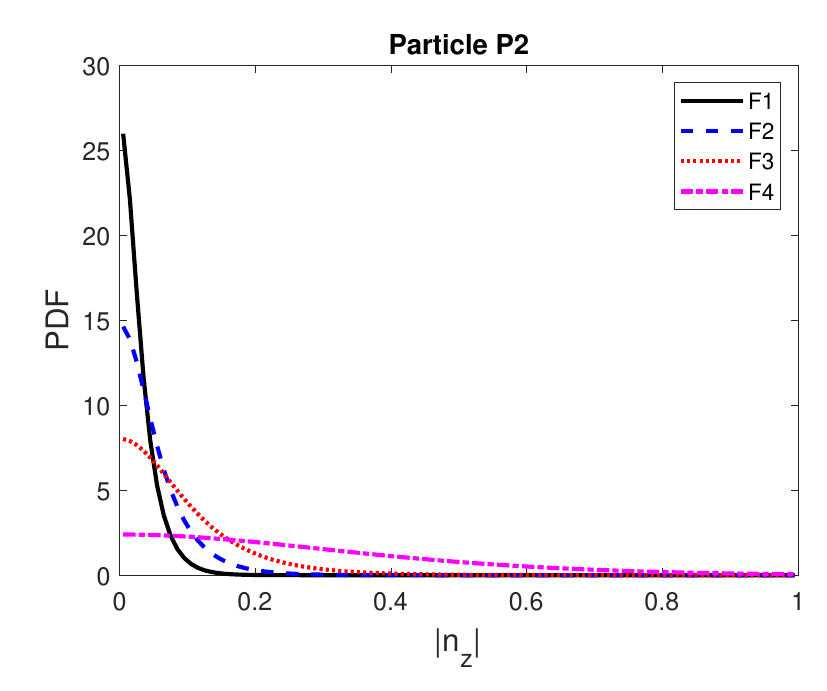}
    \caption{Probability distribution function of the orientation of the settling crystals, represented by the vertical component of $\hat{\mathbf{n}}$, for particle P2 and the four flows considered.
}
	\label{fig:orientation_PDF}
\end{figure}

\begin{figure}[b!]
    \centering
    \includegraphics[trim=1cm 9cm 2cm 9cm,clip,height=5.9cm,keepaspectratio]{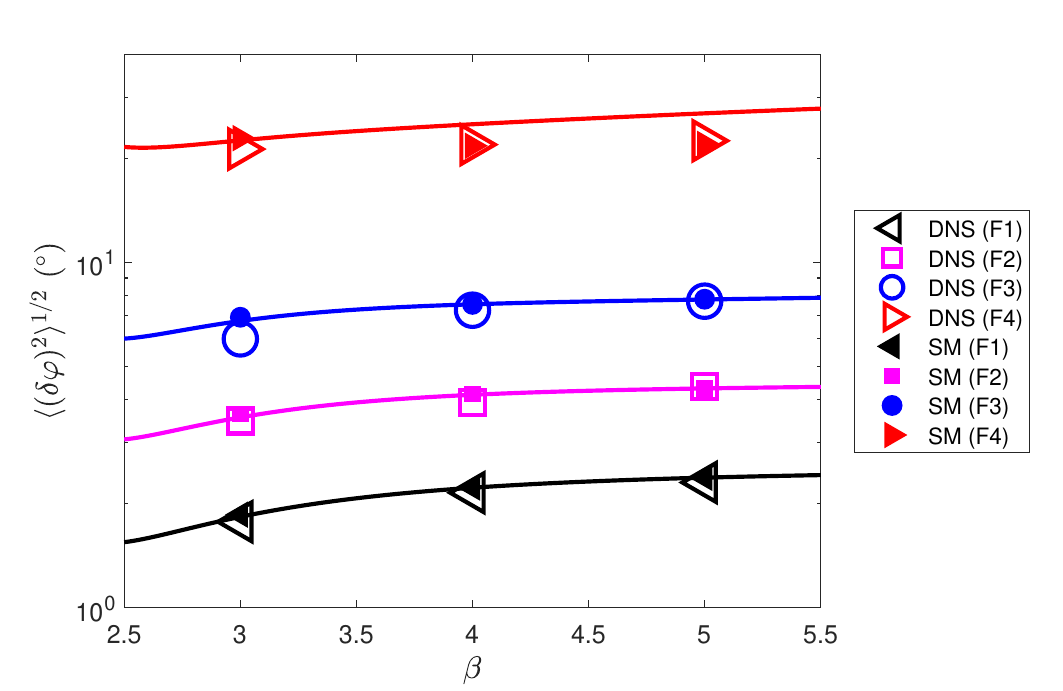}
    \caption{Fluctuations of $\delta\varphi$, the angle of the settling spheroids with respect to the horizontal position ($\delta\varphi=0$ in still fluid), plotted as a function of $\beta$, for flows F1, F2, F3 and F4. Large empty symbols: DNS; small filled symbols: predictions from statistical-model simulations.
    Solid lines show the theory based on the solution~(\ref{eq:dphiODEsolution}) in Appendix~\ref{app:theory}.
}
	\label{fig:orientation_phi}
\end{figure}

\subsection{Particle orientation with respect to gravity} \label{subsec:res_orientation}

We first investigate the statistical orientation of the particles in the presence of gravity. 
We recall here that the hydrodynamic torque acting on these objects, Eq. (\ref{eq:torque_sum}), is the sum of a Stokes contribution and of a correction due to fluid inertia. In a quiescent fluid, any orientation of a particle subject to the first term only is marginally stable, whereas the rotational dynamics of a particle subject to the second term only has a single stable fixed point, corresponding to a horizontal orientation (maximal drag). Spheroidal particles settling in a turbulent flow can therefore have a biased horizontal orientation or a random one, depending on the ratio between the two contributions.
In particular, this subsection extends the results presented in our earlier
work, devoted to disk-shaped particles \cite{She20,Gus21,She22}.

The orientation of an axisymmetric object can be characterized by $n_z$, the projection on the vertical direction $\hat{\bf e}_z=-{\bf g}/\gr$ of the orientation vector $\hat{\mathbf{n}}$, as illustrated in Fig. \ref{fig:n}. Due to the particle fore-and-aft symmetry, two vectors $\hat{\mathbf{n}}$ and $-\hat{\mathbf{n}}$ define the same orientation. Without loss of generality, we will therefore characterize the particle orientation using the absolute value of $n_z$. The probability distribution function (PDF) of $|n_z|$ for particle P2 in the four flows considered is shown in Fig. \ref{fig:orientation_PDF}. Such a distribution would be uniform if all orientations were equally probable. In contrast, the PDFs displayed in Fig. \ref{fig:orientation_PDF} are peaked close to
the $0$ value, which corresponds to a horizontal orientation. 
Interestingly, the distribution is quite sharp in the case of the least 
turbulent flow, F1, 
but becomes broader when the turbulence intensity increases. 
It is worth noticing that 
the orientation of the particle with respect to gravity is sensitive
to the fluid velocity gradient tensor (see ~\cite{Gus21} and 
Sec.~\ref{app:theory} below). The velocity gradient tensor becomes 
increasingly intermittent
when the Reynolds number increases~\cite{Fri97,Buaria:25} (in 
the atmosphere, $R_\lambda$ is much higher than in the simulations corresponding
to Fig.~\ref{fig:orientation_PDF}).
Explicit calculations based on 
an extension of the stochastic model discussed below allow us
to vary the intermittency of the velocity gradient tensor~\cite{Meibohm:2024},
and lead to the conclusion 
that the prediction of the mean value of the distribution is largely 
unaffected by intermittency, as expected theoretically, but that the 
distributions shown 
in Fig.~\ref{fig:orientation_PDF} become somewhat more peaked
when the Reynolds number 
increases. This leads to the conclusion that the distributions observable
for atmospheric turbulence should be somewhat more peaked close to
$0$ than those
shown in Fig.~\ref{fig:orientation_PDF}. However, the distribution
cannot be infinitely peaked around $0$, since the mean value of $n_z$
appears to depend on $\dissip$ only, not on the intermittency of the flow.
This is in qualitative 
agreement with field observations~\cite{Garrett:2015,Grazioli:2025}.
The effect of the particle type, not shown here, is much weaker. Particles therefore settle with a very biased horizontal orientation in flow F1, and their orientation distribution becomes broader at increasing turbulence intensity. This confirms qualitatively our earlier predictions \cite{She20,Gus21}. More precisely, it was shown in \cite{She20} that the orientation bias can be predicted by evaluating the $\mathcal{R}$ parameter, defined as $|{\bf u}-{\bf v}|^2/(\nu|\boldsymbol{\Omega}-\boldsymbol{\omega}|)$, which quantifies the ratio between the inertial torque $\bm{\tau}_h^{(1)}$ and Jeffery's one $\bm{\tau}_h^{(0)}$. For particle P2, the $\mathcal{R}$ parameter varies from $\approx 350$ in F1 to $\approx 25$ in F4, where we have accounted for the factor $C_\tau$ (Eq. (\ref{eq:torque_sum})). The PDFs displayed in Fig. \ref{fig:orientation_PDF} are therefore compatible with those of \cite{She20} (Fig. 3 therein). \\

The particle orientation can be alternatively characterized by the tilt angle $\delta\varphi$, defined as the angle between $\hat{\mathbf{n}}$ and the equilibrium orientation of the particle in a fluid at rest, horizontal (see Fig. \ref{fig:n}). The root-mean-square of $\delta\varphi$ calculated numerically is plotted in Fig. \ref{fig:orientation_phi} as a function of $\beta$, for the four flows considered (empty symbols). This figure confirms quantitatively that the orientation distribution of the particles is very narrowly peaked around $\delta\varphi=0$ when the turbulence intensity is weak, and that its dependence on the particle type is moderate: the fluctuations of $\delta\varphi$ do not exceed $2.5\;\degree$ in flow F1, and increase up to $\approx 20\;\degree$ in flow F4. For a fixed turbulence intensity, they slightly increase with the particle inertia.
We compared the numerical simulation results with those obtained by
replacing the flow in Eqs.~(\ref{eq:NS_1},\ref{eq:NS_2}) with a stochastic velocity field with single time and length scales matched to the Kolmogorov scales~\cite{Gus16}, see Appendix~\ref{app:SM}. This model successfully predicted the alignment of small settling oblate spheroids~\cite{Gus21}.
As shown in Fig.~\ref{fig:orientation_phi}, the model simulations 
(filled symbols) are in very good agreement with the DNS results 
for all four flows and 
the three prolate particles.
However, the
analytical solutions of the model in the limit of small $\delta\varphi$ of \cite{Gus21} (not shown) do not agree so well with the DNS
results, the less so as the particle Stokes number is larger.
To address this difficulty, we
improved this solution to extend its range of validity (Appendix~\ref{app:theory}).
In Fig.~\ref{fig:orientation_phi}, the predictions of the improved
analytical solution (lines) agree well with the 
DNS results, except for the particles with the largest $\st$ 
(run F4).

\begin{figure}[t!]
    \centering
            \includegraphics[trim=3cm 9cm 4cm 9cm, clip,height=5.9cm,keepaspectratio]{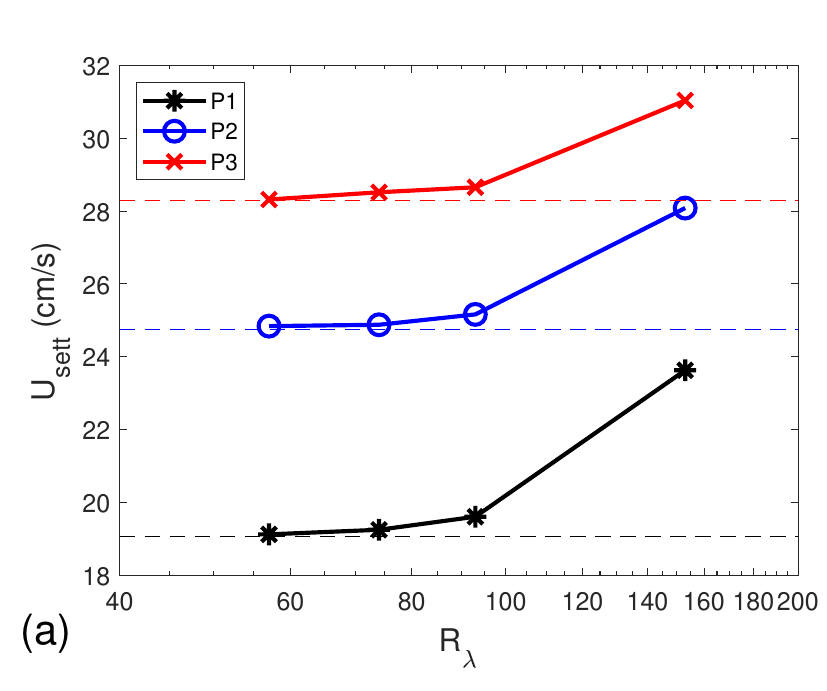}
            \hspace{1.5cm}
            \includegraphics[trim=3cm 9cm 4cm 9cm, clip,height=5.9cm,keepaspectratio]{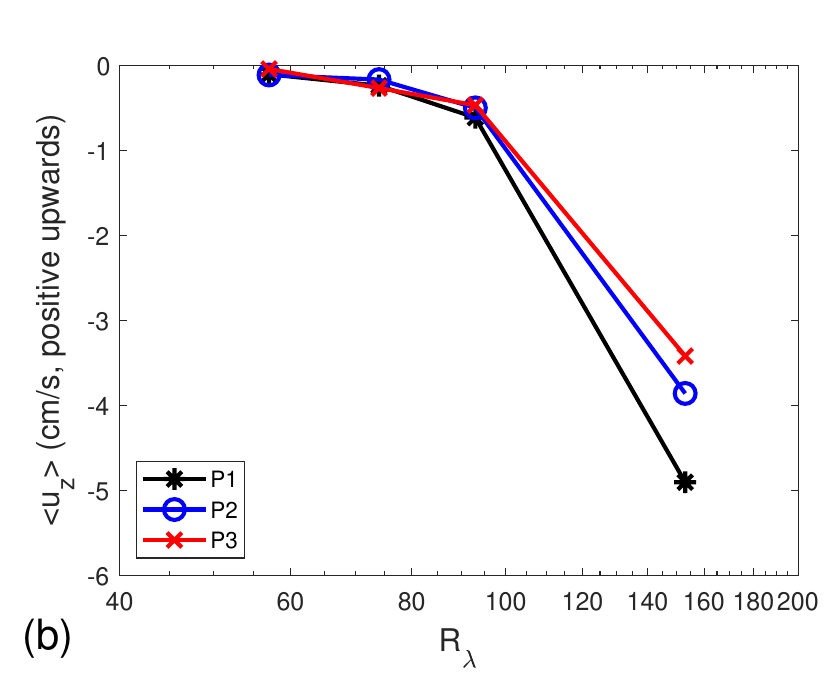}
    \caption{$R_\lambda$ dependence of the averaged: (a) settling velocity of the particle, $U_{\rm sett}$; (b) vertical component of the fluid velocity at the particle position, $\langle u_z \rangle$. In (a) the horizontal dashed lines indicate the values in a fluid at rest, $U_{\rm sett}^{(0)}$, in which the equilibrium orientation is horizontal.}
	\label{fig:eps_vel}
\end{figure}

\subsection{Settling velocity} \label{subsec:sett_vel}

The particle orientation, as well as turbulence, are expected to influence their settling velocity. The averaged settling velocity of the particles, $U_{\rm sett}=|v_z|$, is plotted in Fig. \ref{fig:eps_vel}(a) as a function of the turbulence intensity, for the three particle types. It is compared in the same figure with the settling velocities of the same objects in the same fluid at rest, $U_{\rm sett}^{(0)}$, evaluated using Eq. (A4) from ref. \cite{Bhowmick24}. For a fixed $R_\lambda$, $U_{\rm sett}$ increases with the particle size, from P1 to P3. The effect of turbulence is negligible in flow F1, in which the values of $U_{\rm sett}$ and $U_{\rm sett}^{(0)}$ are virtually indistinguishable. For each particle type, the settling velocity increases with the turbulence intensity. The relative increase of $U_{\rm sett}$ from flows F1 to F4 is $\approx 19\;\%$, $13\;\%$ and $10\;\%$ for particles P1, P2 and P3 respectively.\\

\begin{figure}[t!]
    \centering
            \includegraphics[trim=3cm 9cm 4cm 9cm,clip,height=5.9cm,keepaspectratio]{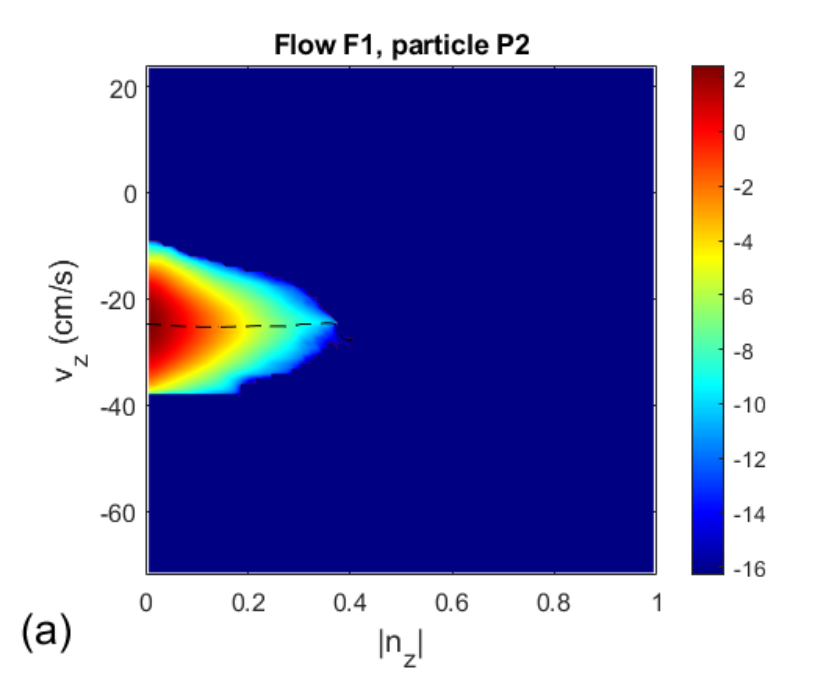}
            \hspace{0.9cm}
            \includegraphics[trim=3cm 9cm 4cm 9cm,clip,height=5.9cm,keepaspectratio]{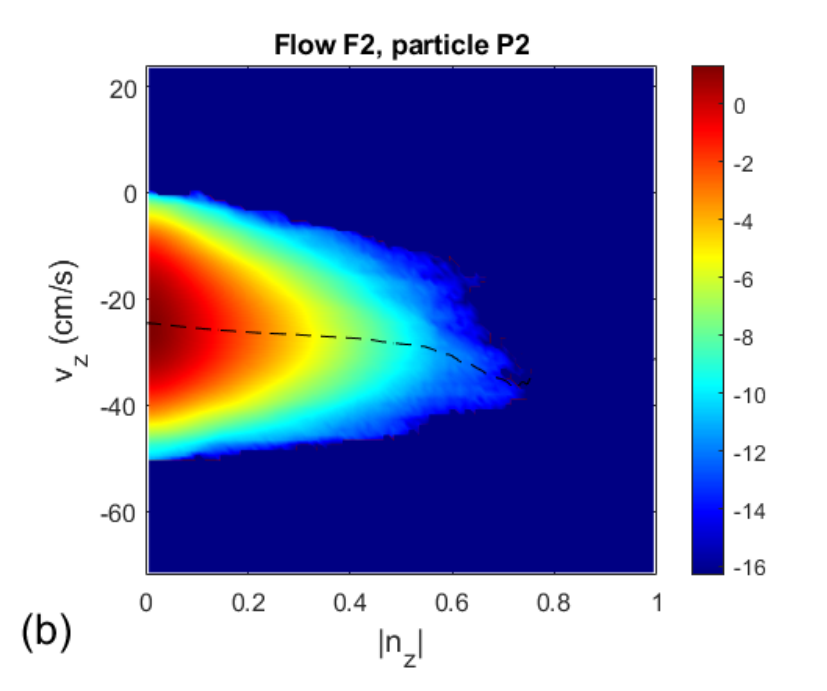}\\
            \vspace{0.5cm}
            \includegraphics[trim=3cm 9cm 4cm 9cm,clip,height=5.9cm,keepaspectratio]{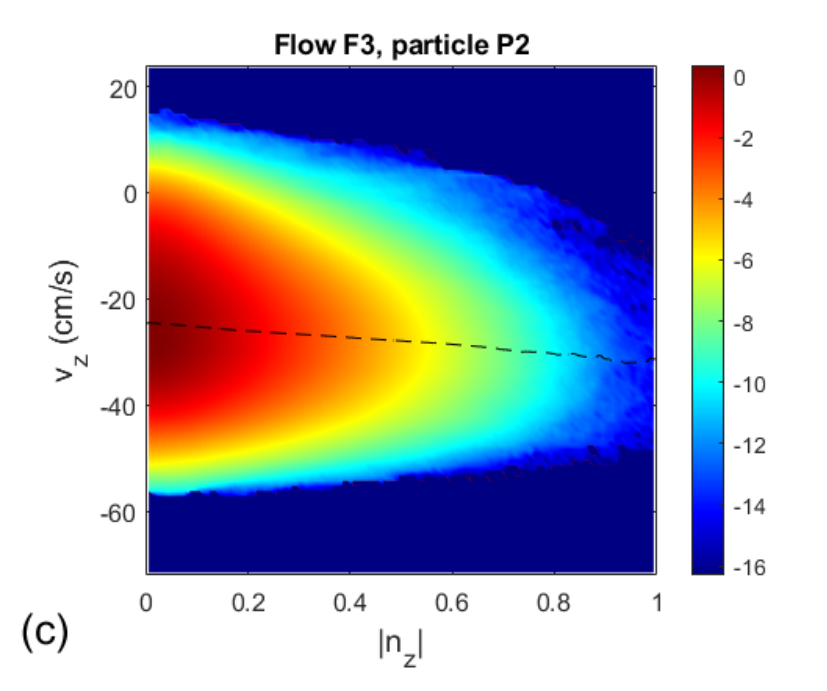}
            \hspace{0.9cm}
            \includegraphics[trim=3cm 9cm 4cm 9cm,clip,height=5.9cm,keepaspectratio]{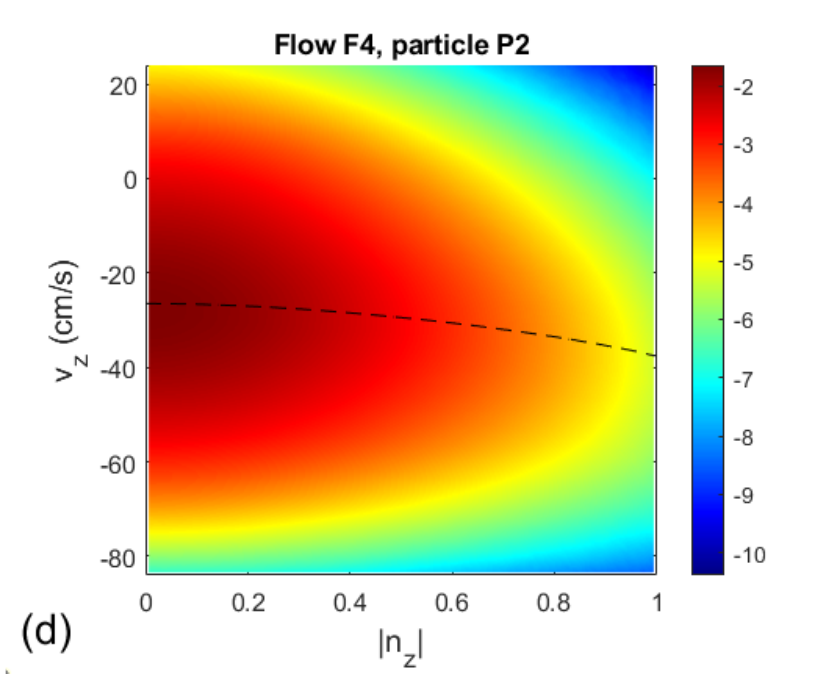}
    \caption{Joint probability distribution function (logarithmic scale) of the particle orientation $|n_z|$ and of the vertical component of the particle velocity, for particle P2: (a) flow F1, (b) flow F2, (c) flow F3, (d) flow F4. The black dashed lines indicate the mean vertical component of the particle velocity conditioned on $|n_z|$.}
	\label{fig:vel_cond}
\end{figure}

To explain this behavior, we recall that the settling velocity of particles in a fluid depends on the way they sample the flow. It was shown in particular that heavy spherical particles preferentially reside in regions of downward flow, which tends to increase their settling velocity~\cite{Max87,Good14,Tom19}. The same behavior was more recently observed for oblate spheroidal particles \cite{She24}. The average vertical component of the fluid velocity at the positions of our prolate objects, $\langle u_z \rangle$, is plotted in Fig. \ref{fig:eps_vel}(b) as a function of the turbulence intensity. For the lowest turbulence intensity considered, this quantity is close to zero. It decreases at increasing $R_\lambda$. In flow F4, for each particle type $- \langle u_z \rangle$ is of the same order as the difference between the settling velocity in this flow and in a fluid at rest. The increase of $U_{\rm sett}$ by turbulence therefore seems to be essentially due to the fact that particles preferentially sample downflow regions of the fluid.
We notice in this respect that our particles are smaller than 
the Kolmogorov scale in the flow ($c \le 0.2\;{\rm mm}$ is at most equal to 
half of the Kolmogorov scale $\eta$, see Table~\ref{tab:flows}) and that the 
particle Reynolds numbers based on the slip velocity are of order $1$.
Judging from results obtained for sedimenting spheres in turbulence, 
one would expect that loitering, as much as
nonlinear corrections
to the drag
should play a role~at higher values of $\st$~\cite{Peng:2025,Ferran23}.

\begin{figure}[t!]
    \centering
            \includegraphics[trim=3cm 9cm 4cm 9cm,clip,height=5.9cm,keepaspectratio]{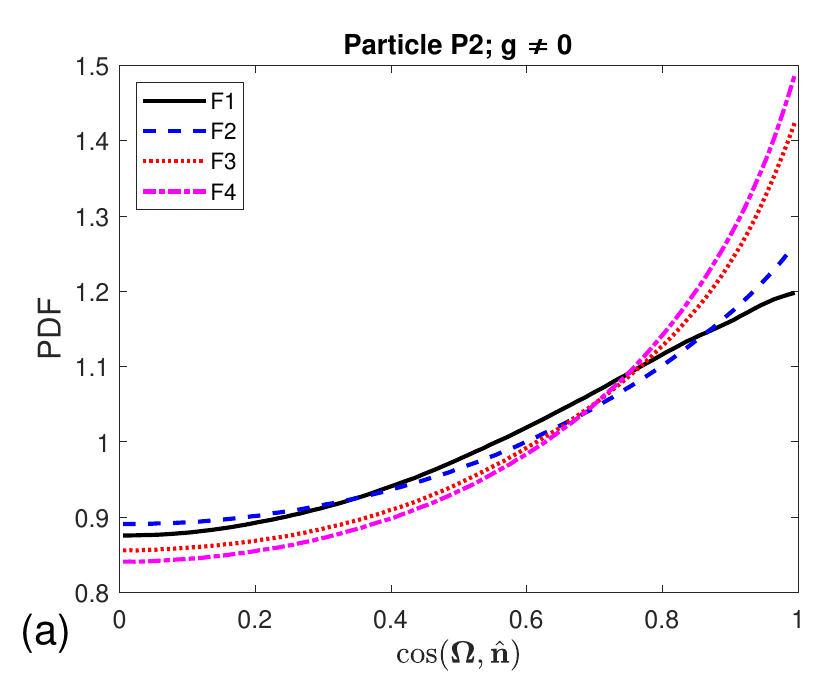}
            \hspace{1.5cm}
            \includegraphics[trim=3cm 9cm 4cm 9cm,clip,height=5.9cm,keepaspectratio]{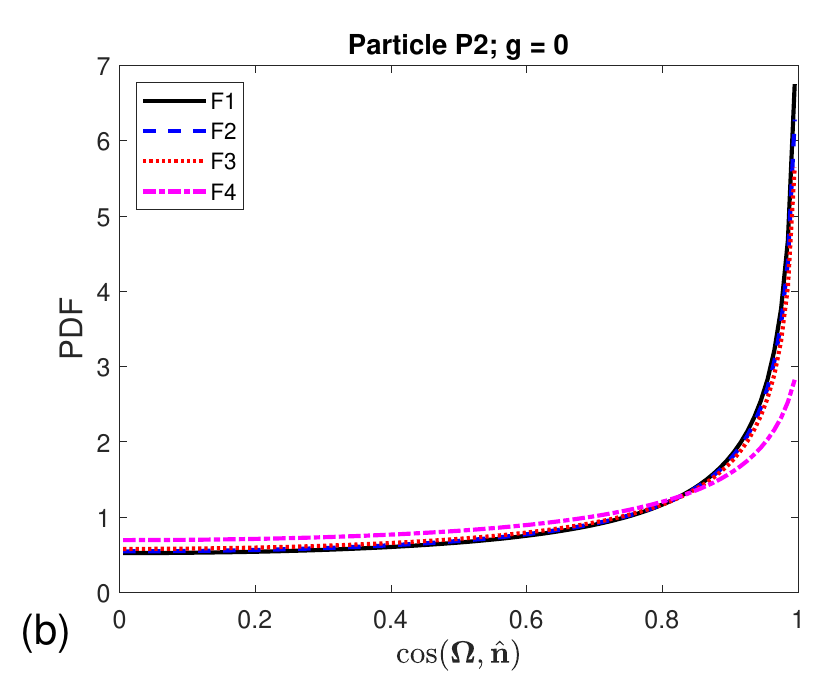}\\
            \vspace{0.5cm}
            \includegraphics[trim=3cm 9cm 4cm 9cm, clip,height=5.9cm,keepaspectratio]{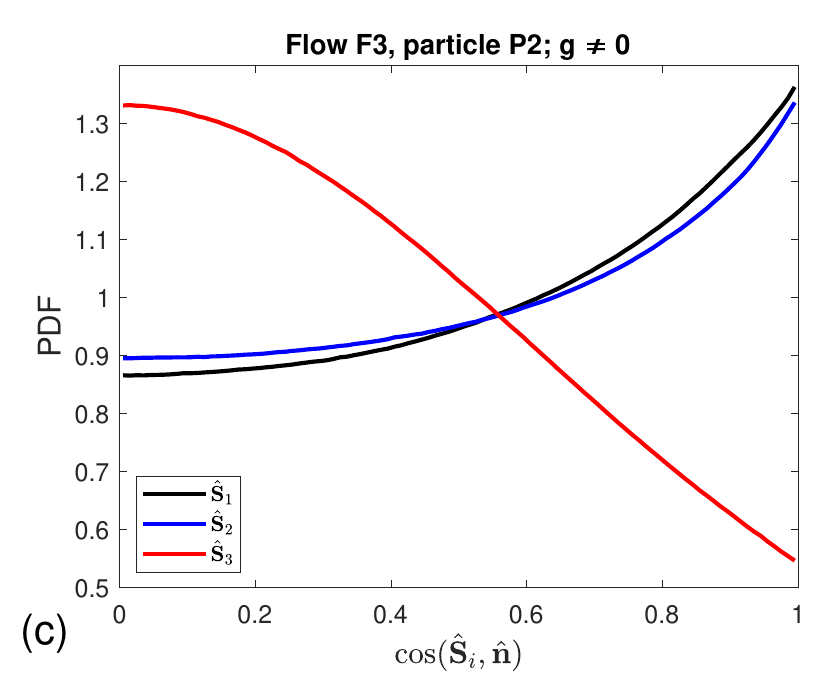}
            \hspace{1.5cm}
            \includegraphics[trim=3cm 9cm 4cm 9cm, clip,height=5.9cm,keepaspectratio]{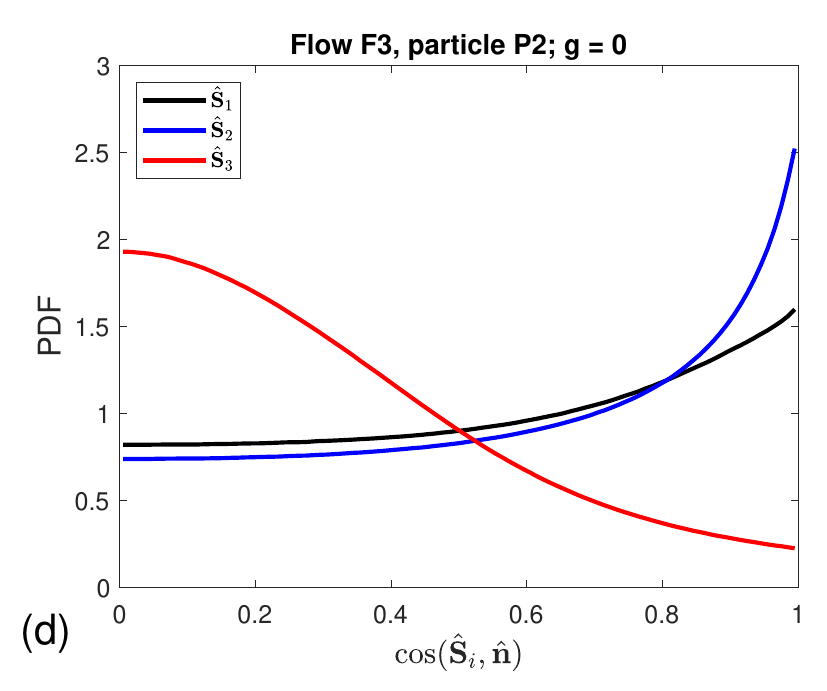}
    \caption{Probability distribution function of the cosine of the angle between the particle orientation $\hat{\mathbf{n}}$ and (a, b) vorticity (flows F1 to F4), (c, d) the strain eigenvectors (flow F3), for particle P2. The statistics have been obtained (a, c) in the presence and (b, d) in the absence of gravity.
}
	\label{fig:om_S_n_PDF}
\end{figure}

It was recently shown that the enhancement of the settling velocity of oblate particles by turbulence can also be explained by their orientation properties \cite{She22,She24}. In a quiescent fluid, an anisotropic object indeed settles in its steady state with a horizontal orientation, which corresponds to a maximal drag, whereas it can explore all other orientations and thereby undergo a lower drag in a turbulent flow. We have evidenced in Sec. \ref{subsec:res_orientation} the same phenomenon for prolate objects. In order to investigate the effect of the orientation of prolate objects on their settling velocity, we have plotted in Fig. \ref{fig:vel_cond} the joint PDF of the vertical component of the particle velocity $v_z$ and of its orientation $|n_z|$, for particle P2 in flows F1 to F4. The vertical velocity $v_z$ conditioned on $|n_z|$ is plotted in dashed lines. Naturally, $v_z$ is always preferentially negative, which reflects the fact that the particles settle on average, and the absolute value of the conditional average $\langle v_z| \ |n_z| \, \rangle$ increases with $|n_z|$: a particle settling vertically ($|n_z|=1$) is subject to a lower drag than the same object oriented horizontally ($n_z=0$). The ratio between $\langle v_z| \ |n_z| \, \rangle (|n_z|=1)$ and $\langle v_z| \ |n_z| \, \rangle (n_z=0)$ is close to $0.79$ in F3 and $0.70$ in F4: as a comparison, this ratio is $\approx 0.78$, and therefore of the same order of magnitude, in a quiescent fluid. At low turbulence intensity, the particle does not explore the more vertical orientations ($|n_z|$ is always $\lesssim 0.4$ in flow F1 and $\lesssim 0.7$ in F2), in agreement with Sec. \ref{subsec:res_orientation}. The moderate value of the ratio $\langle v_z| \ |n_z| \, \rangle (|n_z|=1) / \langle v_z| \ |n_z| \, \rangle (n_z=0)$ added to the fact that the orientation of the particles considered in the present investigation is always biased explains the fact that this orientation does not play a significant role on the settling enhancement by turbulence, as compared to the preferential sweeping effect.

\subsection{Particle alignment with the fluid velocity gradients} \label{subsec:alignt_A}

We now investigate the statistical orientation of particles with respect to vorticity and to the strain eigendirections. We recall here that the real-symmetric strain tensor ${\mathbb S}$ has three real eigenvalues $s_1\ge s_2\ge s_3$ and three orthonormal eigenvectors $\hat{\bf S}_i$ ($i\in \{ 1,2,3 \} $) such that ${\mathbb S}\hat{\bf S}_i=s_i\hat{\bf S}_i$. By incompressibility, $s_1+s_2+s_3=0$, therefore $s_1 \ge 0$ and $s_3 \le 0$. In 3D homogeneous and isotropic turbulence, $s_2$ is on average positive.\\

\begin{figure}[t!]
    \centering
            \includegraphics[trim=3cm 9cm 3cm 9cm,clip,height=5.9cm,keepaspectratio]{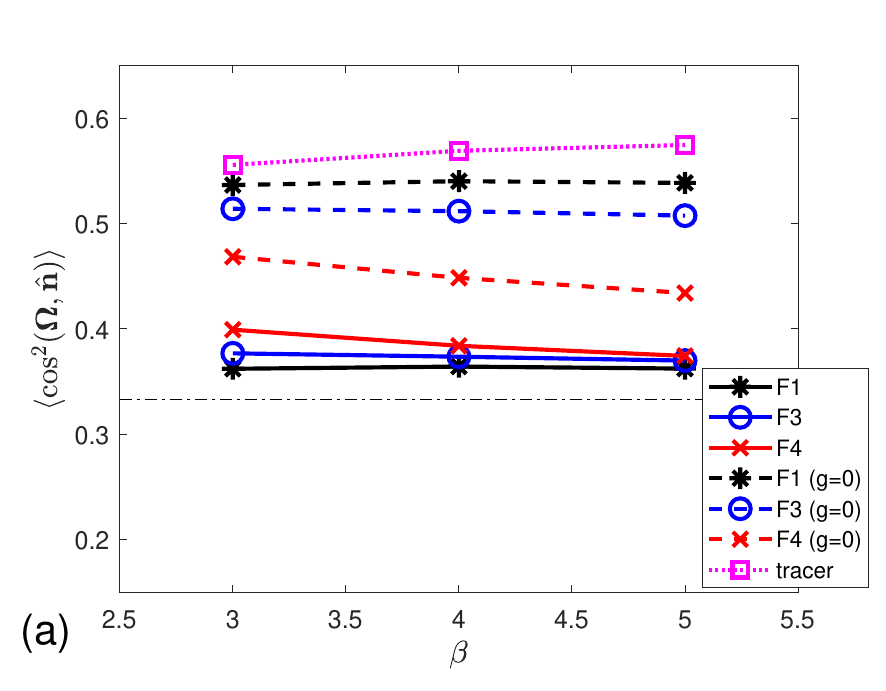}
            \hspace{1.5cm}
            \includegraphics[trim=3cm 9cm 4cm 9cm,clip,height=5.9cm,keepaspectratio]{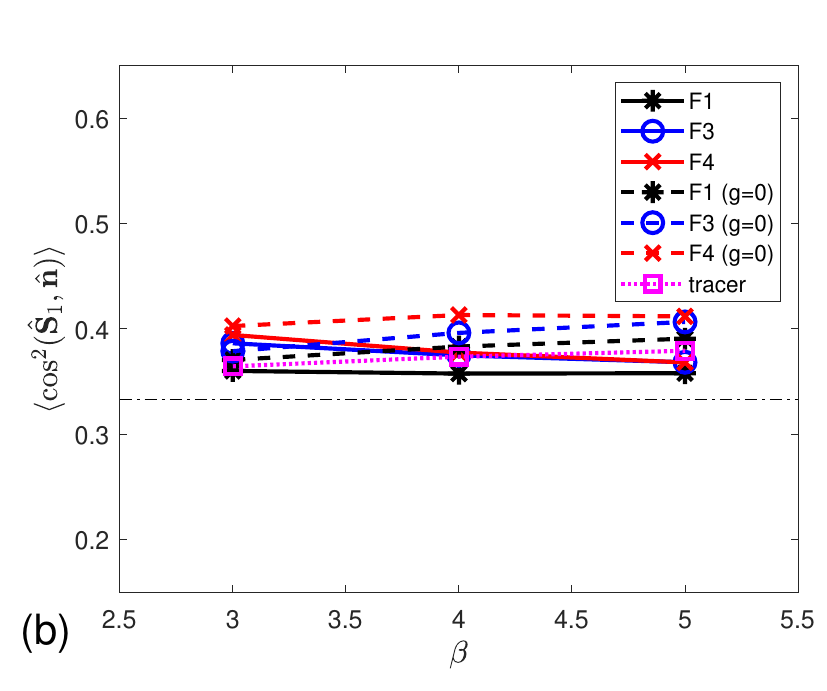}\\
            \vspace{0.5cm}
            \includegraphics[trim=3cm 9cm 3cm 9cm,clip,height=5.9cm,keepaspectratio]{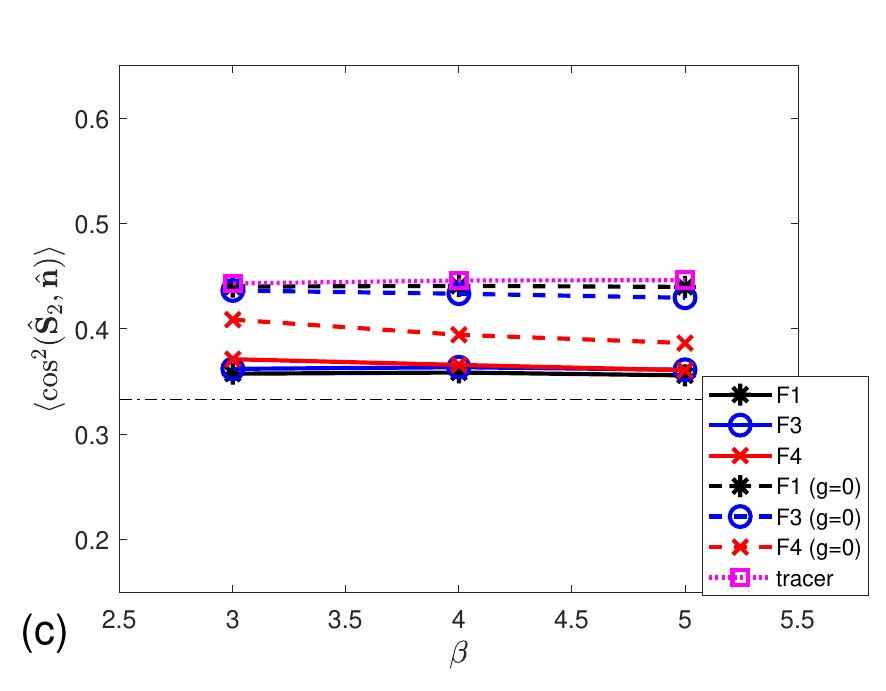}
            \hspace{1.5cm}
            \includegraphics[trim=3cm 9cm 4cm 9cm,clip,height=5.9cm,keepaspectratio]{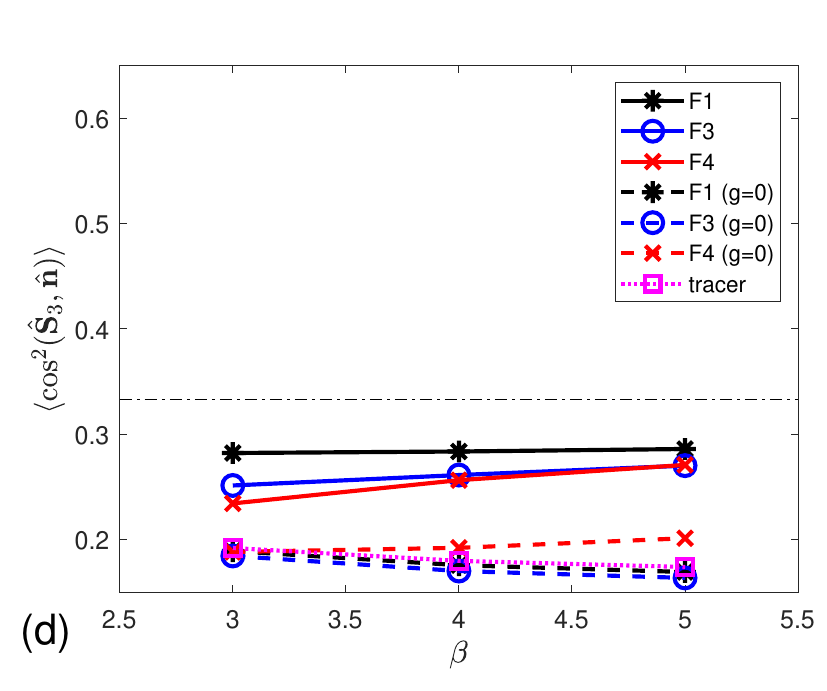}
    \caption{Mean square cosine of the angle between the particle orientation $\hat{\mathbf{n}}$ and (a) vorticity, (b) $\hat{\bf S}_1$, (c) $\hat{\bf S}_2$ and (d) $\hat{\bf S}_3$, plotted as a function of the particle aspect ratio $\beta$. Solid lines: with gravity; dashed lines: without gravity; dotted line: tracer; horizontal dashed-dotted line: value $1/3$ corresponding to the case where all orientations are equally probable.}
	\label{fig:cos2_om_S_n}
\end{figure}

The PDFs of the cosine of the angle between $\hat{\mathbf{n}}$ and these four vectors are displayed for particle P2, with and without gravity, in Fig. \ref{fig:om_S_n_PDF}. The mean square cosine of the same angles is plotted in Fig. \ref{fig:cos2_om_S_n} as a function of the particle aspect ratio, for the different flows considered. In the presence of gravity (Fig. \ref{fig:om_S_n_PDF}(a, c) and solid lines in Fig. \ref{fig:cos2_om_S_n}), the major axis of the particles is preferentially collinear with vorticity $\boldsymbol{\Omega}$, the principal strain axis $\hat{\bf S}_1$, and the intermediate one $\hat{\bf S}_2$, and tends to be normal to $\hat{\bf S}_3$. These tendencies are nevertheless weak. The same trends are exhibited more markedly in the absence of gravity (Fig. \ref{fig:om_S_n_PDF}(b, d) and dashed lines in Fig. \ref{fig:cos2_om_S_n}).\\

In this latter case ($g = 0$), the statistics can be compared to those obtained for tracer particles, {\it i.e.}, microscopic spheroidal objects the center of which is simply advected by the flow, and the rotation of which is subject to Jeffery's torque. We have integrated numerically the equations of motion of such objects with aspect ratios $3$, $4$ and $5$, in flows F1 to F4. The statistical orientation of these objects with the flow was found to be independent of the turbulence intensity, in agreement with results previously obtained in the slender body limit ($\beta\rightarrow + \infty$) \cite{Pum11,Voth16}. The mean square cosine of the angles between $\hat{\mathbf{n}}$ and the velocity gradients for these microscopic objects are plotted in dotted lines in Fig. \ref{fig:cos2_om_S_n}. For all the aspect ratios considered, the tracer particles are preferentially collinear with vorticity, to a lesser extent to $\hat{\bf S}_2$, and even less so to $\hat{\bf S}_1$, and are preferentially normal to $\hat{\bf S}_3$. These trends are qualitatively similar to those obtained for slender bodies \cite{Pum11}. 
In the absence of gravity, the statistical orientation of the inertial particles
with the flow tends to be similar to those of tracers, especially when the 
turbulence intensity decreases (Fig. \ref{fig:cos2_om_S_n}). Such a behavior 
likely reflects the fact that the Stokes numbers of the particles decrease 
with $R_\lambda$ (see Tab. \ref{tab:runs}). Moreover, when $g=0$ the 
preferential alignment of the particles with the velocity gradients overall 
decrease at increasing turbulence intensity, the opposite trend occurring only 
for $\hat{\bf S}_1$. Our own data with particles of the same aspect
ratios and flows with the same energy dissipation $\dissip$ as considered in this 
study
suggest that the dependence of the alignment
between $\nnhat$ and the fluid velocity gradients is a function of $\st$ only (not shown here).\\

In the presence of gravity, the inertial particles are statistically slightly 
more aligned with vorticity and more perpendicular to $\hat{\bf S}_3$ when 
turbulence is more intense (Fig. \ref{fig:om_S_n_PDF}(a) and solid lines in 
Fig. \ref{fig:cos2_om_S_n}(a, d)). This could be explained by the fact 
that the horizontal orientation of the objects due to the inertial torque is 
then reduced: the importance of Jeffery's torque increases with 
$R_\lambda$, therefore restoring the alignments between $\nnhat$ and velocity 
gradient directions, at least at moderate $\st$. 
In other words, as expected, the influence of gravity decreases when 
increasing turbulence intensity (Fig. \ref{fig:cos2_om_S_n}).
When $\st$ becomes large enough, however, the alignment effect diminishes. 
This explains qualitatively why for flow F4, a lighter particle is better
aligned with vorticity and more perpendicular to $\hat{\bf S}_3$ than a heavier one (solid lines in Fig. \ref{fig:cos2_om_S_n}(a, d)).

\begin{figure}[b!]
    \centering
            \includegraphics[trim=4.5cm 9cm 5cm 9cm, clip,height=5.3cm,keepaspectratio]{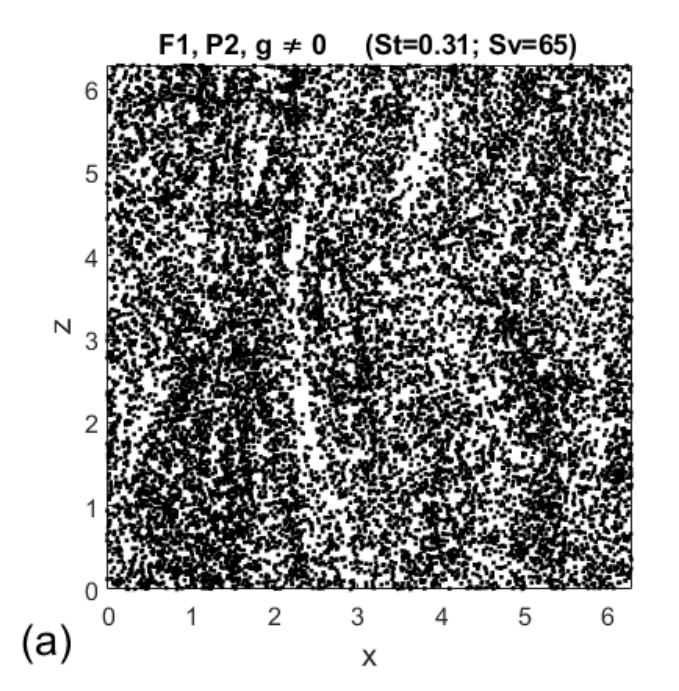}
            \hspace{0.1cm}
            \includegraphics[trim=4.5cm 9cm 5cm 9cm, clip,height=5.3cm,keepaspectratio]{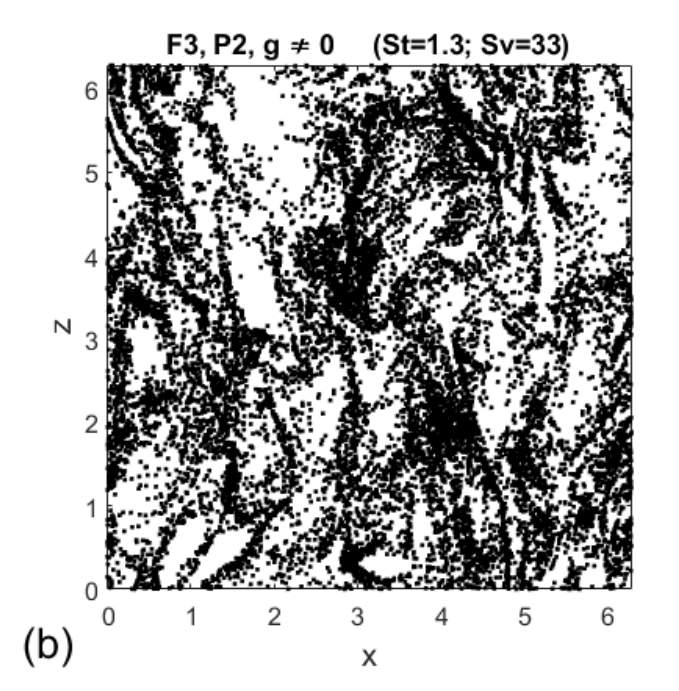}
            \hspace{0.1cm}
            \includegraphics[trim=4.5cm 9cm 5cm 9cm, clip,height=5.3cm,keepaspectratio]{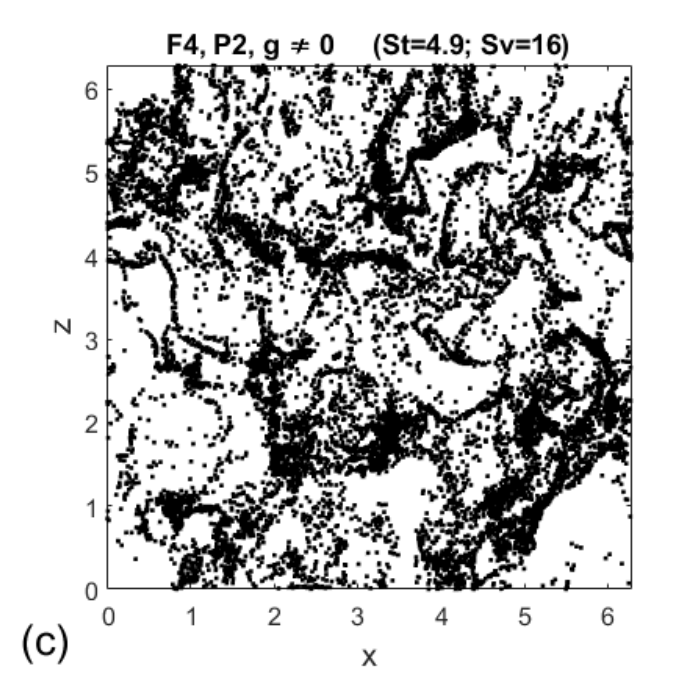}
            \\
            \includegraphics[trim=4.5cm 9cm 5cm 9cm, clip,height=5.3cm,keepaspectratio]{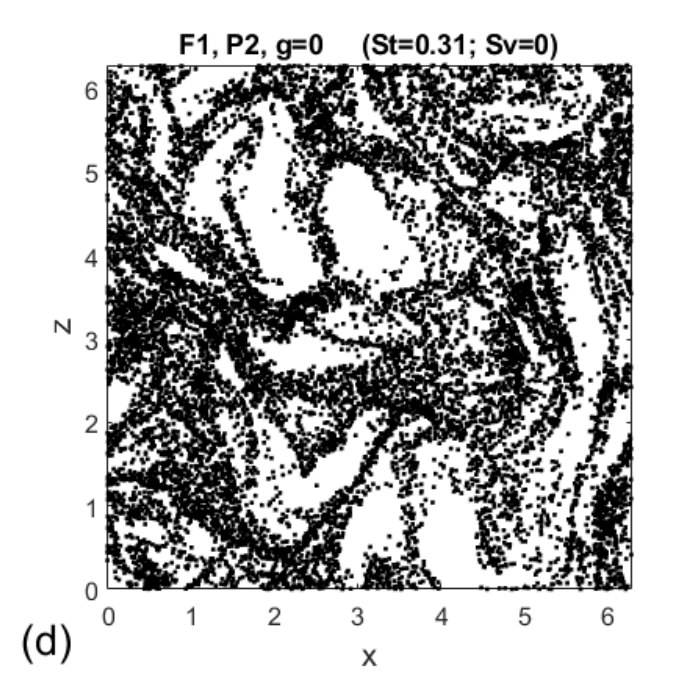}
            \hspace{0.1cm}
            \includegraphics[trim=4.5cm 9cm 5cm 9cm, clip,height=5.3cm,keepaspectratio]{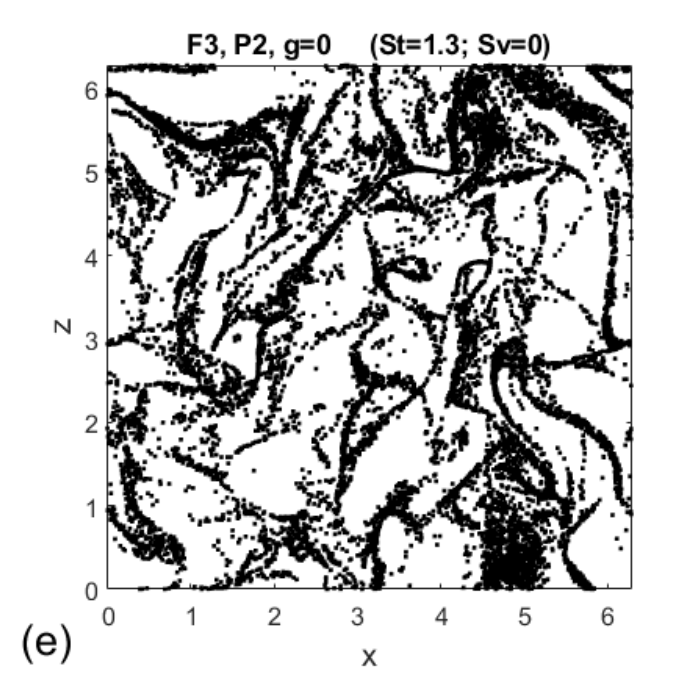}
            \hspace{0.1cm}
            \includegraphics[trim=4.5cm 9cm 5cm 9cm, clip,height=5.3cm,keepaspectratio]{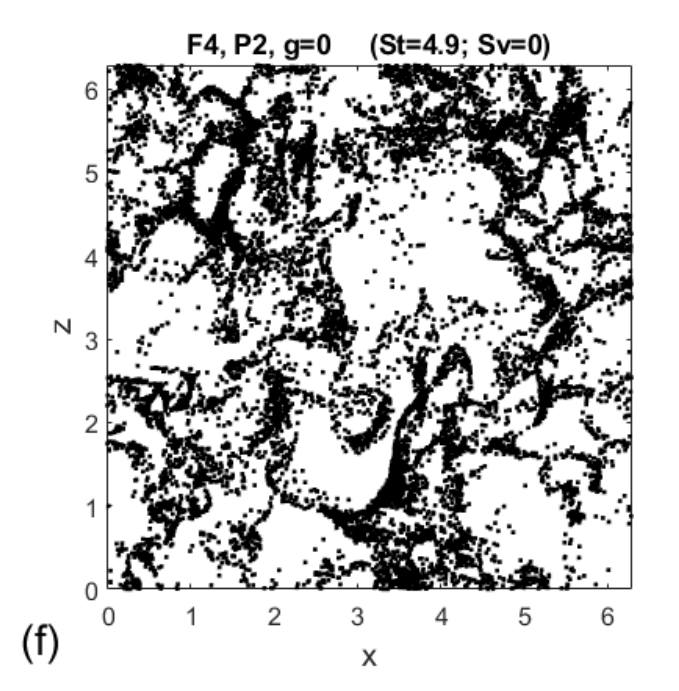}
    \caption{Typical instantaneous distributions of the particles P2 in a vertical plane $(x,z)$, of thickness $0.016\,L_{\rm box}$ in the $y-$direction: (a, d) flow F1, (b, e) flow F3, (c, f) flow F4. (a, b, c) With gravity; (d, e, f) without gravity.}
	\label{fig:part_distr}
\end{figure}

\subsection{Spatial distribution of particles in the flow} \label{subsec:distribution}

We finally investigate the particles distribution in the flow. Figure \ref{fig:part_distr} displays typical instantaneous distributions in a vertical plane, for particle P2. These distributions are overall strongly inhomogeneous, with very pronounced voids and clusters. This is especially true when gravity is switched off (bottom line of Fig. \ref{fig:part_distr}). In the presence of gravity, the distribution heterogeneity is maximal in flow F4 (Fig. \ref{fig:part_distr}(c)). In flow F1, the particles are distributed more evenly, although vertical structures can be distinguished (Fig. \ref{fig:part_distr}(a)). We note that vertical structures have been clearly observed for settling spheres in DNS of turbulent flows~\cite{Woittiez09,Flor20}, as well as in simulations of stochastic models of turbulent flows~\cite{Gus14a}.

\begin{figure}[t!]
    \centering
            \includegraphics[trim=3cm 9cm 4cm 9cm,clip,height=5.9cm,keepaspectratio]{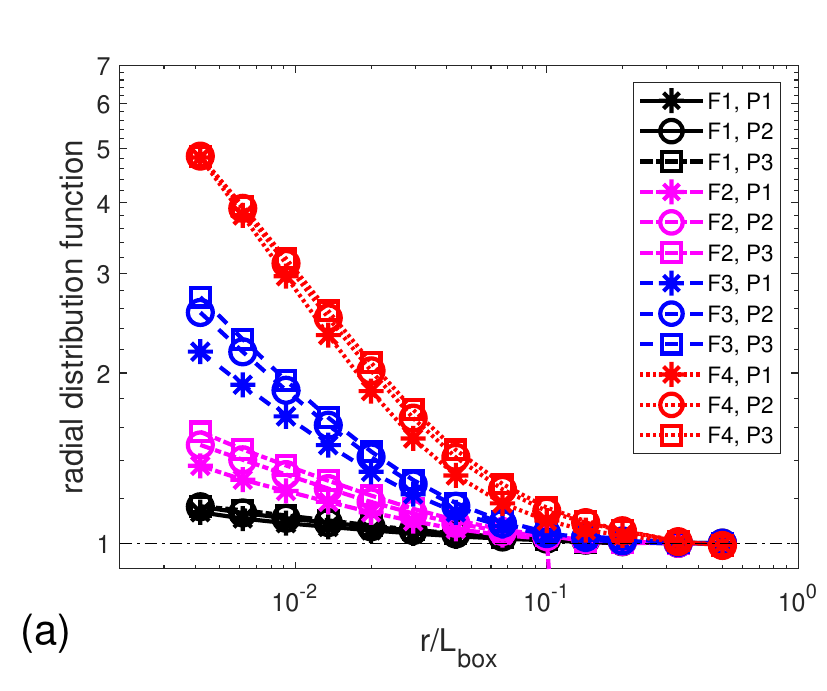}
            \hspace{1.5cm}
            \includegraphics[trim=3cm 9cm 4cm 9cm,clip,height=5.9cm,keepaspectratio]{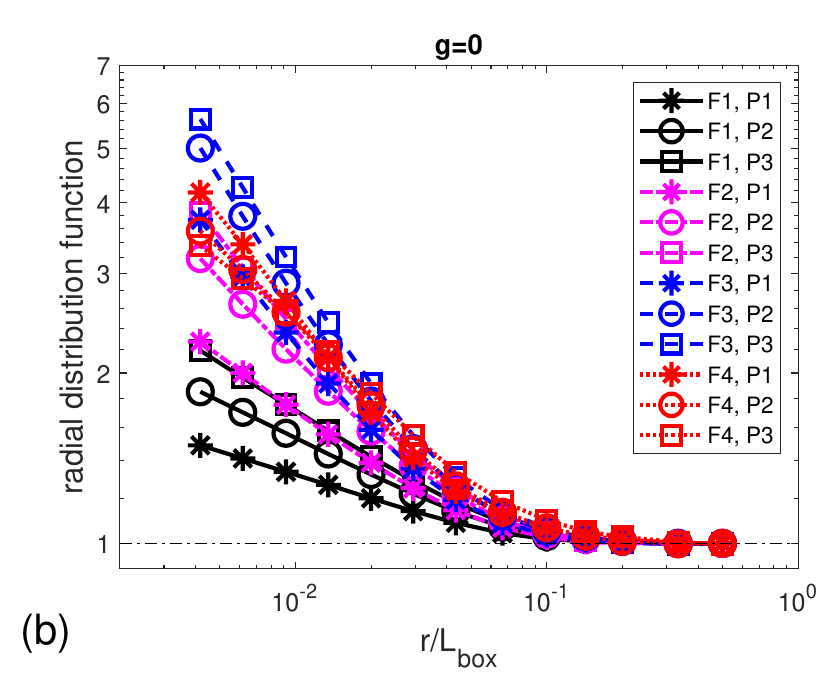}\\
            \vspace{0.5cm}
            \includegraphics[trim=3cm 9cm 4cm 9cm,clip,height=5.9cm,keepaspectratio]{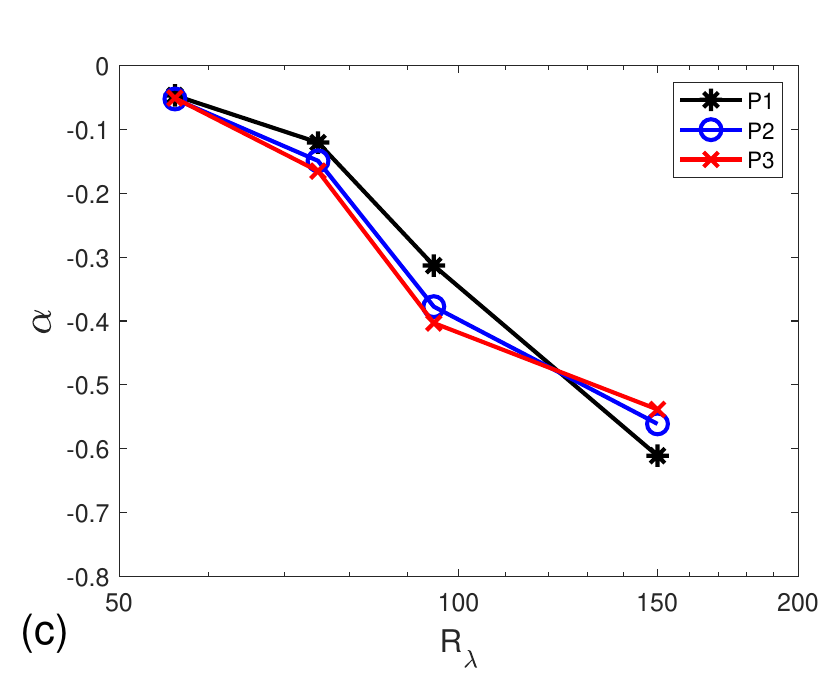}
            \hspace{1.5cm}
            \includegraphics[trim=3cm 9cm 4cm 9cm,clip,height=5.9cm,keepaspectratio]{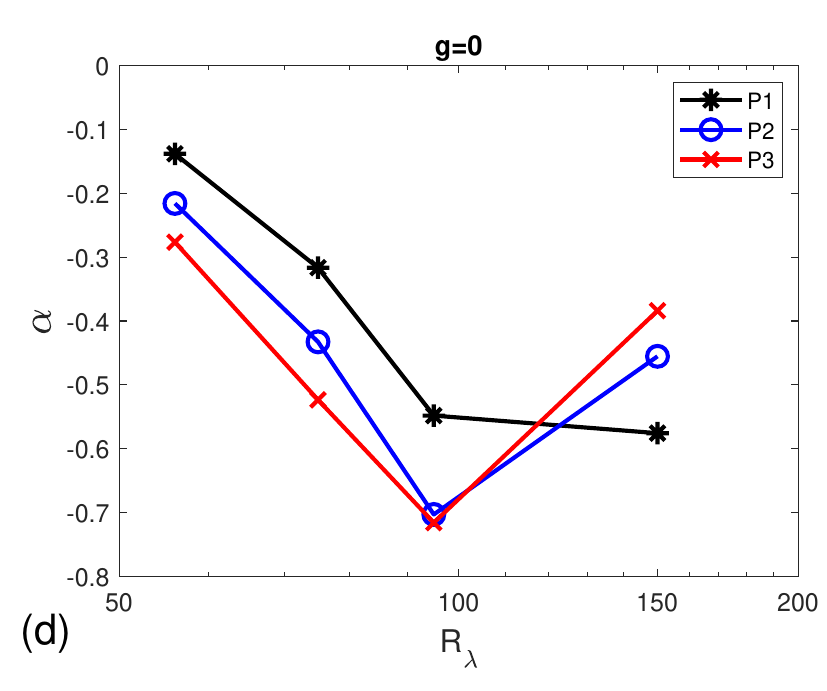}
    \caption{(a, b) Radial distribution function plotted as a function of the normalized distance $r/L_{\rm box}$; (c, d) $R_\lambda$-dependence of the exponent $\alpha$ measured at short distance $r$ (first five markers in panels (a) and (b)), for the three particles. (a, c) With gravity; (b, d) without gravity.
}
	\label{fig:rdf}
\end{figure}

The particle distribution can be characterized more quantitatively by calculating the three-dimensional radial distribution function (rdf)~\cite{McQuarrie_book,HOLTZER_COLLINS_2002,Bec:24}.
This function is defined for a statistically isotropic system of identical particles as the ratio of the number of particle pairs found at a given separation distance $r$ to the expected number if the particles were uniformly distributed. The rdf is plotted in Fig. \ref{fig:rdf}(a,b) for all the flows and particles considered, both in the presence and in the absence of gravity. The value $1$ is reached only for distances $r\gtrsim L_{\rm box}/10$. For smaller distances, the rdf is larger than $1$, thereby indicating that the probability of finding a pair separated by $r$ is higher than it would be for a homogeneous distribution. This is a clear signature of the clusters displayed in Fig. \ref{fig:part_distr}. For a given value of $r$ and in the presence of gravity (Fig. \ref{fig:rdf}(a)), the deviation of the rdf from $1$ at small scales increases monotonously with the turbulence intensity, and depends much more weakly on the particle type. In the absence of gravity, see Fig. \ref{fig:rdf}(b), the dependence of the rdf on $R_\lambda$ is non monotonic: the highest values of the radial distribution function are then reached for flow F3.\\

\begin{figure}[t!]
    \centering
            \includegraphics[trim=3cm 9cm 4cm 9cm,clip,height=6.3cm,keepaspectratio]{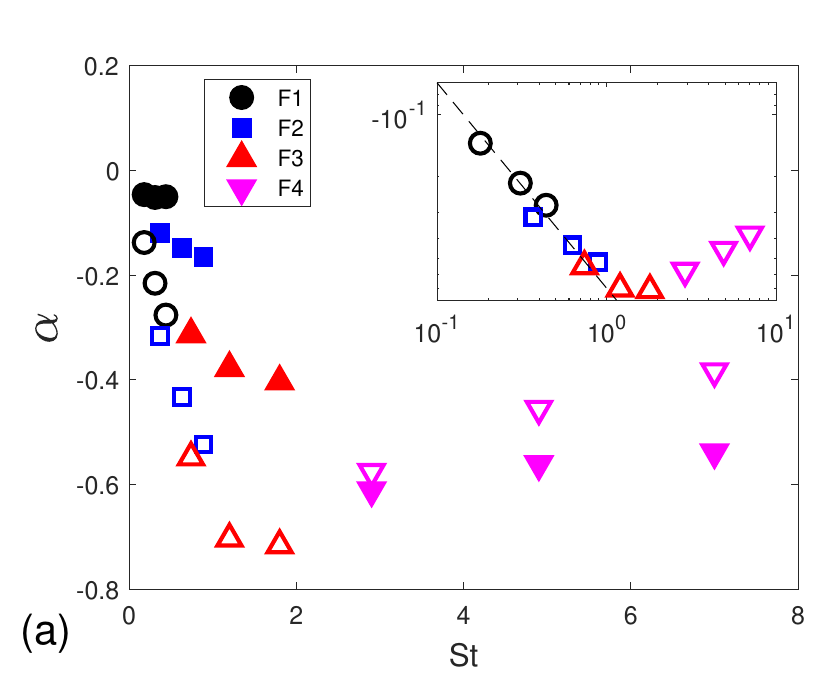}
            \hspace{0.5cm}
            \includegraphics[trim=3cm 9cm 4cm 9cm,clip,height=6.3cm,keepaspectratio]{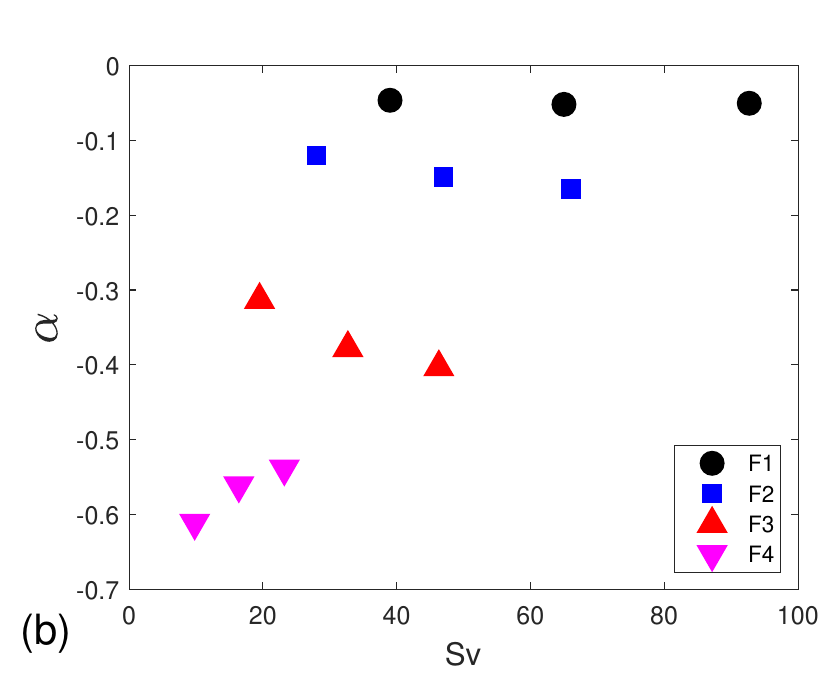}
    \caption{ Dependence of the exponent $\alpha$ of the radial distribution 
function measured at small scale on $\st$ (a) and $\sv$ (b). Filled (resp. empty) symbols correspond to simulations with (resp. without) gravity. In (a), 
the exponents are plotted in the inset using logarithmic scales 
(in the absence of gravity only). The dashed line indicates a 
$\sim \st^{-1}$ dependency of $\alpha$.
}
	\label{fig:rdf_St_Sv}
\end{figure}

At small values of $r$, the rdf displays a power-law behavior, related to the correlation dimension of the fractal set characterizing the particle spatial distribution~\cite{Bec:24}. The corresponding exponent $\alpha$ (equal to $D_2-3$, where $D_2$ is the correlation dimension in the dynamical systems framework~\cite{Bec:24}) provides a quantitative measurement of the inhomogeneity of the particles distribution. 
The exponents $\alpha$ are measured by fitting the curves shown in Fig. \ref{fig:rdf}(a,b) (first five markers) by a power law; their dependence 
on $R_\lambda$ is shown in  Fig. \ref{fig:rdf}(c,d), which 
confirms quantitatively the trends visible in Fig. \ref{fig:rdf}(a,b). 
Namely, we observe that the effects of preferential concentration increase ({\it i.e.}, $|\alpha|$ increases)
with $R_\lambda$ in the presence of gravity, as suggested by 
Fig.~\ref{fig:part_distr}(a) and (c). On the other hand, without gravity, 
the dependence on $R_\lambda$ is not monotonous, which confirms the 
observations of Fig.~\ref{fig:part_distr}(d-f).  Note however that the 
increase of $R_\lambda$ is due to the larger values of $\dissip$ from flow F1 to F4, hence of $\st$ and $\sv$. The analysis can be taken further by plotting $\alpha$ as a function of the dimensionless numbers $\st$ and $\sv$, see Fig. \ref{fig:rdf_St_Sv}. Remarkably, in the absence of gravity, the exponent $\alpha$ appears to have a simple functional dependence on $\st$ (see the empty symbols in Fig. \ref{fig:rdf_St_Sv}(a)), suggesting that the particle distribution depends only on this dimensionless parameter, independently of other particle characteristics such as dimensions, aspect ratio, etc, and flow properties, at least over the range of parameters covered in this study. 
Remarkably, we observe a linear dependence of $\alpha$ on $1/\st$ for $\st \lesssim 1$, as shown in the inset of Fig.~\ref{fig:rdf_St_Sv}(a). This observation is generally consistent with what has been observed for spheres, although theoretical predictions in that case rather suggest a $\st^{-2}$-dependence~\cite{Bec:24}. As already known for spherical objects \cite{Bec07}, the preferential concentration effect is the strongest for $\st\sim 1$. For small values of $\st$ the particles follow the flow, whereas they have their own trajectory at large $\st$: in both situations the particles are expected to be distributed evenly in the fluid.
To take the comparison between rods and spheres a step further, we have used the stochastic model described in Appendix \ref{app:SM} to calculate the rdf for spherical particles of sizes allowing to match the settling velocities of the spheroids. The rdf thus obtained for spherical particles do not deviate much from those of the spheroids (themselves consistent with our DNS), which means that the degree of clustering of the rods considered in the present study is to a large degree described by the clustering of spheres.
Finally, it is important to stress that 
the anisotropy affecting 
the motion of the particles
remains relatively weak. 
Using the notation 
introduced in Eq.~\eqref{eq:def_Mt}, we find that
$A_\|/A_\perp \approx 1.23$ for $\beta = 3$ and 
$\approx 1.33$ for $\beta = 5$ (the ratio is $1$ for spheres). For
this reason, the
similarity to the clustering of spheres may be lower for
particles with a much larger aspect ratio.\\

In the presence of gravity, the rdf exponent is expected to depend both on $\st$ and $\sv$. Its absolute value is also maximal for a value of $\st$ close to $1$ (Fig. \ref{fig:rdf_St_Sv}(a)).We observe that $\alpha$ is generally larger in the presence of gravity, at least for $\st \lesssim 2$. This is due to the weaker effect of turbulence, as the settling tends to decorrelate the action of the flow on the preferential concentration. Note however that for $\st \gtrsim 2$, the effect is opposite. This could be attributed to the multiplicative amplification discussed in~\cite{Gus14a}. The dependence on $\sv$ is more intricate, although an increasing $\sv$ roughly induces a more homogeneous distribution of the particles reflected by a smaller value of $| \alpha |$ (Fig. \ref{fig:rdf_St_Sv}(b)). 
This observation can be interpreted by invoking the crossing trajectory effect \cite{Csanady63}, according to which the rapid travel of heavy particles across the turbulent eddies can result in an appreciable reduction of their dispersion.
Figure~\ref{fig:rdf_St_Sv}(b) clearly points to a strong dependence on the flow, hence on the Froude number.

\section{Conclusions}
\label{sec:conclusion}

In this work, we studied the statistical properties of columnar
ice crystals settling through a turbulent flow, in conditions representative
of turbulent clouds. Our work rests on a few simplifying
approximations: we assume that turbulence is homogeneous
and isotropic, therefore ignoring the complex structure of the flow in a cloud,
in particular near the edges.
In view of the small sizes of the crystals, we also modeled the forces and
torques acting on the crystals by using some simplified expressions, which
have been thoroughly derived~\cite{Dab15} and checked numerically~\cite{Froh20,Ouch20,jiang2021inertial} and validated experimentally in some special cases
(e.g. in the
absence of turbulence~\cite{Roy:2019,Cabrera22,Bhowmick24}).
Although the size of the computational domain, $ L_{\rm box} \sim 25 \, {\rm cm}$,
is much smaller than the large integral scales relevant to clouds, the values
of the turbulent energy dissipation $\dissip$, are realistic.

This study extends our own work focusing on thin plate-like crystals
settling in turbulent flows~\cite{She22}. An important difference with this
work is that the inertia of the columnar crystals studied here is
significantly larger than that of the plate-like crystals considered before.
We are also focusing on a range of values of the aspect ratio which is far less
extreme than the one considered before.

Our study confirms the general trend that anisotropic crystals tend to
settle with their broadest sides facing down. Turbulence acts to broaden
the distribution of angles $\delta\varphi$, as introduced in Fig.~\ref{fig:n}, the
more so as $\dissip$ increases. Quantitatively, the numerical results
are very accurately reproduced by the statistical model of~\cite{Bec:24},
consistent with results obtained in the case of oblate crystals~\cite{Gus21,She22}.

Our results point to a significant increase of the averaged settling
velocity when the turbulent energy dissipation rate increases. Although
one could expect that the broader distribution of the angle of the particle
with respect to the horizontal, $\delta\varphi$, may contribute to the enhancement of
the settling velocity when $\dissip$ increases, our numerical results
rather suggest that the effect of alignment of $\hat{\mathbf{n}}$ perpendicular
to the direction of gravity is weak, at least for the crystals considered
($3 \le \lambda \le 5$).
Instead, our numerical results point to a very
strong preferential sampling effect:
particles tend to accumulate in regions where the fluid velocity points
downwards~\cite{Max87}. The effect is particularly strong at the highest
value of $\dissip$ considered here, for which the Stokes numbers of the
particles are in the range $ 2.9 \le \st \le 7.0 $, {\it i.e.} when the inertia
effects of the particles become very significant.

Although the Stokes number of the particles is moderate at weak
turbulence intensity, $0.18 \le \st \le 0.44$ in the case of
flow F1 with $\dissip = 4.43\; {\rm cm}^2/{\rm s}^3$, one observes very
significant clustering effects. These can be clearly seen by visualizing the
particle distribution (see Fig.~\ref{fig:part_distr}).  The radial
distribution function exhibits a classical power law dependence at
small distances, the exponent providing a quantitative characterization of
the clustering. The clustering effect diminishes when the settling parameter
$\sv$
increases, which can be understood since the effect of turbulence is
decorrelated by the vertical motion of the particles. We notice in this
respect that, in absence of any gravity ($\sv = 0$), the dependence of the
clustering for columnar crystals is qualitatively similar as in the case
of droplets (spheres).

Finally, we also documented the properties of alignment of the axis
of the particle, $\hat{\mathbf{n}}$, with the vorticity $\bm{\Omega}$
and with the eigenvalues of the rate of
strain. The preferential alignment between
$\hat{\mathbf{n}}$ and $\bm{\Omega}$ in the absence of inertia ($\st = 0$)
and of  gravity ($ \sv = 0$)~\cite{Pum11} is disrupted by settling,
as $\hat{\mathbf{n}}$ tends to be preferentially perpendicular to
$\hat{\mathbf{e}_z}$. Still, we do observe a weak preferential alignment
of $\hat{\mathbf{n}}$ with $\bm{\Omega}$, significantly less than in the
absence
of gravity. Similarly, we observe a weak alignment of $\hat{\mathbf{n}}$ with
the intermediate eigenvalue of strain.

The results of the present study are expected to be important when studying
collisions between columnar crystals, in the spirit of~\cite{She22}. One of
the interesting differences with the case of disk-like crystals considered
in~\cite{She22} is the very weak dependence of the settling velocity
on the angle between $\hat{\mathbf{n}}$ and the vertical direction. This
implies that the collision mechanism based on differential settling should not
play any significantly role for the columnar crystals considered. On the other
hand, the much stronger values of the Stokes numbers for the
columnar crystals considered should favor collision mechanisms based on the
sling effect~\cite{She22}. A detailed study of collisions between settling
columnar crystals will be the subject of future work.

\acknowledgments
We acknowledge the PSMN
and the datacenter at the Ecole Normale Sup\'erieure de Lyon
for computational resources.

\appendix

\section{Expressions of tensors and functions involved in the equations of motion of a spheroid} \label{app:appendix}

We provide here explicit expressions of the tensors and functions involved in the equations of motion of a spheroidal particle settling in a fluid.\\

The expression of the moment of inertia tensor of a spheroid, ${\mathbb I}$, reads:
\begin{equation}
I_{ij} = I_\perp (\delta_{ij} - n_i n_j) + I_\| n_i n_j, ~~~~ {\rm with } ~~~~
I_\perp = \frac{1 + \beta^2}{5} a^2 ~~~ {\rm and} ~~~ I_\| = \frac{2}{5} a^2.
\label{eq:mom_inert}
\end{equation}

The translational resistance tensor ${\mathbb A}$ (Eqs. (\ref{eq:f0}) and (\ref{eq:drag_correction})) is equal to \cite{Kim:2005}:
\begin{equation}
A_{ij} \equiv
A_\perp (\delta_{ij}-n_i n_j) + A_\parallel n_i n_j\,,
\label{eq:def_Mt}
\end{equation}
with  coefficients
\begin{align}
A_\perp&=\frac{8(\beta^2-1)}{3\beta[(2\beta^2-3)\gamma+1]}\,,\quad
A_\parallel=\frac{4(\beta^2-1)}{3\beta[(2\beta^2-1)\gamma-1]}\,,\quad
\gamma=\frac{\ln[\beta + \sqrt{\beta^2-1}]}{\beta\sqrt{\beta^2-1}}\,.
\nonumber
\end{align}
These expressions are consistent with those given in Tables~3.4 and 3.6 in \cite{Kim:2005}.\\

The rotational resistance tensors ${\mathbb C}$ and ${\mathbb H}$ in Eq.(\ref{eq:tau0}) have
the following explicit expressions:
\begin{align}
C_{ij} &\equiv
C_\perp (\delta_{ij}-n_i n_j) + C_\parallel n_i n_j\,,
\quad H_{ijk} =H_0\epsilon_{ijl} n_k n_l\,,
\label{eq:def_Mr}
\end{align}
with
\begin{align}
C_\perp&=\frac{8 a^2 (\beta^4-1)}{9\beta [(2\beta^2-1)\gamma-1]}\,,\quad C_\parallel=-\frac{8 a^2 (\beta^2-1)}{9(\gamma-1)\beta}\,,\quad
H_0= -C_{\perp}\frac{\beta^2 - 1}{\beta^2 + 1}\,.
\nonumber
\end{align}
Here $\epsilon_{ijl}$ is the antisymmetric tensor, and the Einstein summation convention is used.\\

The function $F(\beta)$ in Eq. (\ref{eq:tau1}) is a shape factor computed by \cite{Dab15}. For a prolate spheroid ($\beta > 1$),
\begin{equation}
\begin{split}
F(\beta)=&\frac{-\pi e^2(420e+2240e^3+4249e^5-2152e^7)}{315((e^2+1)\tanh^{-1}e-e)^2((1-3e^2)\tanh^{-1}e-e)}\\
&+\frac{\pi e^2(420+3360e^2+1890e^4-1470e^6)\tanh^{-1}e}{315((e^2+1)\tanh^{-1}e-e)^2((1-3e^2)\tanh^{-1}e-e)}\\
&-\frac{\pi e^2(1260e-1995e^3+2730e^5-1995e^7)(\tanh^{-1}e)^2}{315((e^2+1)\tanh^{-1}e-e)^2((1-3e^2)\tanh^{-1}e-e)},
\end{split}
\label{eq:Fbeta}
\end{equation}
where the ellipticity $e$ is defined as $ e =  \sqrt{1 - (1/\beta)^2 }$.

\section{Stochastic model}
\label{app:SM}
In the stochastic model, the velocity field is expressed in terms of a Gaussian-distributed vector potential, $\ve u(\ve x,t)=\ve\nabla\times\ve A(\ve x,t)/\sqrt{d(d+2)}$, where the components of $\ve A$ have zero mean and covariance~\cite{Gus16}:
\begin{align*}
\langle A_i(\ve x,t)A_{i'}(\ve x',t')\rangle=u_{\rm f}^2\ell_{\rm f}^2\delta_{ii'}\exp\left(-\frac{|\ve x-\ve x'|^2}{2\ell_{\rm f}^2}-\frac{|t-t'|}{\tau_{\rm f}}\right)\,.
\end{align*}
Here $u_{\rm f}=\langle\ve u^2\rangle^{1/2}$ is the root-mean-square velocity, and $\ell_{\rm f}$ and $\tau_{\rm f}$ are the correlation length and time, respectively.
These parameters define the dimensionless Kubo number, $\ku=u_{\rm f}\tau_{\rm f}/\ell_{\rm f}$.

For large $\ku$, the stochastic model can be matched to turbulence by identifying the Kolmogorov time, $\tau_\eta=\langle{\rm tr}(\ma A^{\sf T}\ma A)\rangle^{-1/2}=\ell_{\rm f}/(\sqrt{5}u_{\rm f})$, where $\ma A$ is the flow velocity-gradient matrix, and the Kolmogorov length, $\eta=0.1\ell_{\rm f}$, corresponding to the smooth scale observed in DNS of turbulence~\cite{Gus16,Bec:24}.
In our simulations we set $\ku=10$.

\section{Theory for small tilt angles of settling spheroids}
\label{app:theory}

The dynamics of prolate spheroids is governed by Eqs.~(\ref{eq:Newt_transl}--\ref{eq:inertial_corrections}).
The dominant contribution to the angular dynamics for prolate spheroids, when the tilt angle $\dphi$ is small, was analyzed in Ref.~\cite{Gus21}.
Namely, the dynamics (see Eq.~(18b) in Ref.~\cite{Gus21}) is given by:
\begin{align}
\begin{split}
\tfrac{{\rm d}}{{\rm d}t}\dphi&=\omega_s\,,\quad\tfrac{{\rm d}}{{\rm d}t}\theta=\omega_g\,,\\
\tfrac{{\rm d}}{{\rm d}t}\omega_g &= \tfrac{\Co}{\Io\st}(-\omega_g - Y_{sp}+ \dphi Y_{gs})\,,\\
\tfrac{{\rm d}}{{\rm d}t}\omega_p &=\tfrac{\Cp}{\Ip\st}(-\omega_p + \Omega_p)\,,\\
\tfrac{{\rm d}}{{\rm d}t}\omega_s &=\tfrac{\Co}{\Io\st}(-\omega_s + Y_{gp} - Y_{gg}\dphi)\,.
\end{split}
\label{eq:omega_spherical_smalldphi_columns}
\end{align}
These equations are made dimensionless using the Kolmogorov time $\tau_\eta$ and length $\eta$.
Here it is assumed that $\sv$ is large and that $\st$ is not too large, consistent with the parameters of flows F1, F2, and F3, while the flow F4 lies on the boundary to this regime (Table~\ref{tab:runs}).
Moreover, $\theta$ is the polar angle of $\nnhat$, and the subscripts $g$, $p$, and $s$ denote contractions with the unit vectors $\gghat$, $\pphat=\nnhat+\gghat\tan\dphi$, and $\sshat=\gghat\times\pphat$, respectively.
The elements of the matrix $\ma Y$ in the Cartesian basis are
\begin{align}
Y_{ij} = -C_\tau\tfrac{\Io}{\Co}{\mathscr A}' W_i W_j -B_{ij}\,,
\label{eq:YDef}
\end{align}
where $\ve W$ is the slip velocity, and $\ma B=\ma O+\Lambda\ma S$ with $\ma O$ and $\ma S$ being the anti-symmetric and symmetric parts of the fluid-gradient matrix, respectively.
The parameters $\Cp$ and  $\Co$ are the components of the resistance tensor in Eq.~(\ref{eq:def_Mr}), $\Ip$ and $\Io$ are the moments of inertia in Eq.~(\ref{eq:mom_inert}), and ${\mathscr A}'=\tfrac{5}{6\pi}F(\beta)\tfrac{\beta^3}{1+\beta^2}$, with $F(\beta)$ in Eq.~(\ref{eq:Fbeta}) being negative for prolate spheroids.
Finally, we modified the model in Ref.~\cite{Gus21} by inserting the correction factor $C_\tau$ to the torque due to fluid inertia in Eq.~(\ref{eq:torque_sum}), consistent with~Ref.~\cite{Bhowmick24}.

The corresponding expansion of the translational dynamics, including the correction to the force due to fluid inertia with a correction factor $C_f$, becomes
\begin{align}
\begin{split}
\tfrac{{\rm d}}{{\rm d}t}v_g \!&=\!-\tfrac{\Ao}{\st}W_g\big[1+C_f\tfrac{3a_\perp}{8}\Ao W_g\big] + \tfrac{\sv}{\st}\,,\\
\tfrac{{\rm d}}{{\rm d}t}v_p \!&=\!\tfrac{\sigma}{\st}W_p
+\tfrac{\Delta}{\st}\dphi W_g\,,\,\sigma=-\Ap\big[1+C_f\tfrac{3a_\perp}{16}(3\Ap-\Ao)W_g\big]\,,\,\Delta=(A_\parallel-A_\perp)\big[1+C_f\tfrac{3a_\perp}{16}(2\Ao+3\Ap)W_g\big]\\
\tfrac{{\rm d}}{{\rm d}t}v_s \!&=\!-\tfrac{\Ao}{\st}W_s\big[1+C_f\tfrac{3a_\perp}{8}\Ao W_g\big]\,.
\end{split}
\label{eq:v_spherical_smalldphi_columns}
\end{align}
Here $\Ap$ and $\Ao$ are the components of the resistance tensor in Eq.~(\ref{eq:def_Mt}).
These equations are valid under the same assumptions that were invoked to derive Eq.~(\ref{eq:omega_spherical_smalldphi_columns}).
In this limit, $\sv$ is large and $\dphi$ is small, allowing to approximate the vertical slip velocity by the steady-state speed of horizontally aligned spheroids settling in quiescent flow, giving
\begin{align}
W_g=v_g^*=\tfrac{4}{3C_fa_\perp A_\perp}\Big(\sqrt{1+\tfrac{3}{2}C_fa_\perp \sv}-1)\Big)\hspace{0.5cm}\mbox{and}\hspace{0.5cm}Y_{gg}=-C_\tau\tfrac{\Io}{\Co}{\mathscr A}'[v_g^*]^2\,.
\end{align}

The dynamics of the subsystem $\dphi$, $\omega_s$, and $v_p$ with $v_g=v_g^*$ in Eqs.~(\ref{eq:omega_spherical_smalldphi_columns}) and (\ref{eq:v_spherical_smalldphi_columns}) only couples to the other variables through the trajectory dependence of the flow.
By approximating the trajectories along which the flow is evaluated by $\ve x_t^{(\rm d)}=\ve x_0+v_g^*\gghat t$, the subsystem decouples.
This dynamics is planar: it resides in the $\pphat$-$\sshat$ plane. It was earlier studied in Ref.~\cite{Bhowmick24} for the case of particles settling in quiescent flow.
Here, we explicitly solve this dynamics.
The three coupled first-order differential equations for $\dphi$, $\omega_s$ and $v_p$ can be rewritten as a third-order differential equation for $\dphi$ by applying the operator $e^{\sigma t/\st}\tfrac{{\rm d}}{{\rm d}t}e^{-\sigma t/\st}\tfrac{{\rm d}}{{\rm d}t}$ to the equation $\tfrac{{\rm d}}{{\rm d}t}\dphi-\omega_s=0$, where $\sigma$ is defined in Eq.~(\ref{eq:v_spherical_smalldphi_columns}).
The resulting equation becomes
\begin{align}
\tfrac{{\rm d}^3}{{\rm d}t^3}\dphi+\tfrac{1}{\st}[\tfrac{\Co}{\Io}-\sigma]\tfrac{{\rm d}^2}{{\rm d}t^2}\dphi
+\tfrac{\Co}{\Io\st}[Y_{gg}-\tfrac{\sigma}{\st}]\tfrac{{\rm d}}{{\rm d}t}\dphi
-\tfrac{\Co}{\Io\st^2}(\sigma+\Delta)Y_{gg}\dphi
=\tfrac{\Co}{\Io\st}\big[
\tfrac{1}{\st}\sigma B_{gp}(t)-\tfrac{Y_{gg}}{v_g^*}\tfrac{{\rm d}}{{\rm d}t}u_p(t) - \tfrac{{\rm d}}{{\rm d}t}B_{gp}(t)
\big] .
\label{eq:dphiODE}
\end{align}

Equation~(\ref{eq:dphiODE}) is a driven third-order linear differential equation with constant coefficients.
Its solution is
\begin{align}
\begin{split}
\dphi&=\int_0^t{\rm d}t'\left[\tfrac{e^{r_1t_1}[f(t-t_1)+r_1g(t-t_1)]}{(r_1-r_2)(r_1-r_3)}+\tfrac{e^{r_2t_1}[f(t-t_1)+r_2g(t-t_1)]}{(r_2-r_3)(r_2-r_3)}+\tfrac{e^{r_3t_1}[f(t-t_1)+r_3g(t-t_1)]}{(r_3-r_1)(r_3-r_2)}\right]\,.
\end{split}
\label{eq:dphiODEsolution}
\end{align}
Here $f(t)=\tfrac{C_\perp}{I_\perp}\tfrac{\sigma}{\st^2} B_{gp}(t)$ and $g(t)=-\tfrac{\Co}{\Io\st}\big[\tfrac{Y_{gg}}{v_g^*}u_p(t) + B_{gp}(t)\big]$ are given by the driving terms in Eq.~(\ref{eq:dphiODE}), and $r_1$, $r_2$, and $r_3$ are the roots to the characteristic polynomial of the left-hand side in Eq.~(\ref{eq:dphiODE}), all having negative real part when the fixed point $\dphi^*=0$ of the undriven dynamics is stable. The time is assumed to be large, so the contribution from the initial condition ($\sim e^{r_it}$) is negligible.
Squaring Eq.~(\ref{eq:dphiODEsolution}) and evaluating the average along settling trajectories gives the steady-state average $\langle\dphi^2\rangle$ in the limit $t\to\infty$ in terms of $r_i$, the parameters of the problem, and the correlation functions of $f$ and $g$.
For the stochastic model in Appendix~\ref{app:SM}, these correlation functions are known, allowing for analytical computation of $\langle\dphi^2\rangle$.
Evaluation of this expression with numerical evaluation of the roots $r_i$ gives the theoretical values plotted in Fig.~\ref{fig:orientation_phi} (lines).
Data points for general aspect ratios were obtained by linear interpolation of $c$ and $\rho_{\rm p}$ in Table~\ref{tab:particles}.

\begin{figure}[t!]
    \centering
    \includegraphics[trim=1cm 9cm 2cm 9cm,clip,height=5.9cm,keepaspectratio]{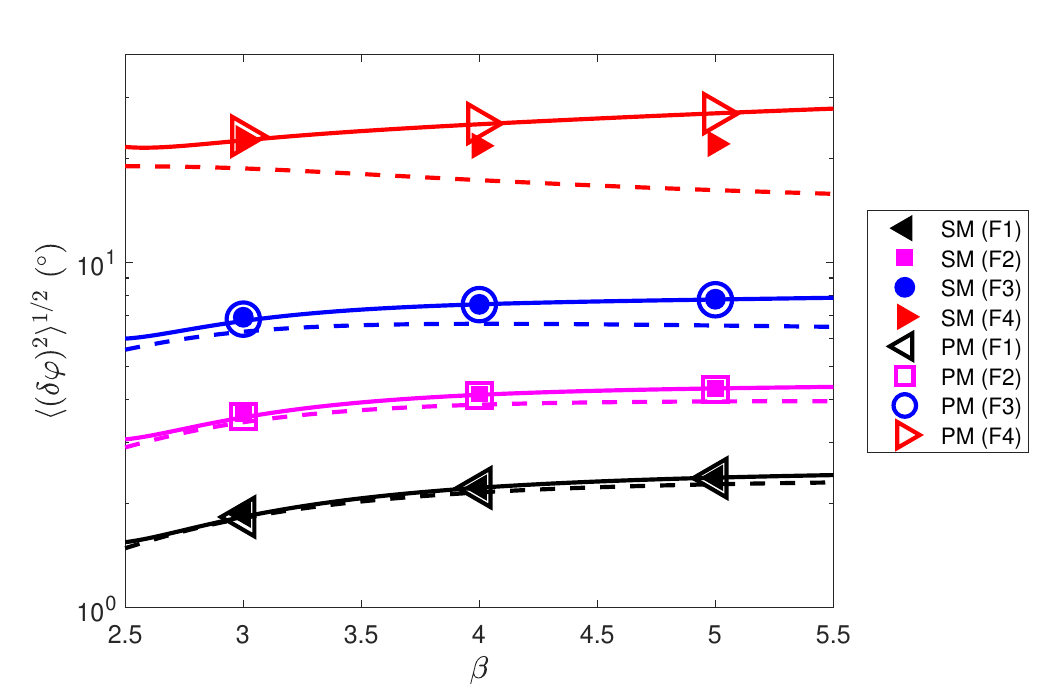}
    \caption{Same as Fig.~\ref{fig:orientation_phi}, but comparing the stochastic model simulations (small, solid markers) to statistical-model simulations of the simplified planar dynamics (large, empty markers) ($\dphi$, $\omega_s$ and $v_p$ from Eqs.~(\ref{eq:omega_spherical_smalldphi_columns}) and (\ref{eq:v_spherical_smalldphi_columns}) evaluated along settling trajectories, $v_g=v_g^*$ and $\ve x=\ve x^{{\rm d}}=\ve x_0+v_g^*\gghat t$).
    Lines compare the theory in Fig.~\ref{fig:orientation_phi} (Eq.~(\ref{eq:dphiODEsolution}), solid lines) to the theory based on the solution to Eq.~(\ref{eq:dphiODEfirst}) (dashed).
}
	\label{fig:orientation_phi_comparison}
\end{figure}

Figure \ref{fig:orientation_phi_comparison} shows simulations of the stochastic model for the full dynamics (solid symbols, same as Fig.~\ref{fig:orientation_phi}) and the planar model (empty symbols).
The results agree well for flows F1, F2 and F3, but do not agree for flow F4.
For all flows, including F4, the simulations of the planar model agree with the theoretical prediction from Eq.~(\ref{eq:dphiODEsolution}) (solid lines), showing that the theory is indeed the exact solution of the planar model.
This implies that the deviation between theory and simulations observed for flow F4 in Fig.~\ref{fig:orientation_phi} are solely because the approximations of the planar model fail when the Stokes number and tilt angles become too large.

\subsection{Relation to earlier theoretical results}
Equation~(\ref{eq:dphiODEsolution}) gives the general solution to the planar model.
In Ref.~\cite{Gus21}, limiting cases of this solution were identified.
Below, we explain how these solutions relate to Eq.~(\ref{eq:dphiODE}).

First, multiplying Eq.~(\ref{eq:dphiODE}) by $\st^2$ and considering the limit of small $\st$ to remove the third -and second-order derivatives of $\dphi$ yields
\begin{align}
\tfrac{\Co}{\Io}[Y_{gg}\st-\sigma]\tfrac{{\rm d}}{{\rm d}t}\dphi
-\tfrac{\Co}{\Io}(\sigma+\Delta)Y_{gg}\dphi
=\tfrac{\Co}{\Io}\Big[
\sigma B_{gp}(t)-\st\tfrac{Y_{gg}}{v_g^*}\tfrac{{\rm d}}{{\rm d}t}u_p(t) - \st\tfrac{{\rm d}}{{\rm d}t}B_{gp}(t)
\Big] .
\label{eq:dphiODEfirst}
\end{align}
Further, assuming $\sv$ is large enough, so $Y_{gg}\st-\sigma\approx Y_{gg}\st$ in the first term, and solving the equation for $\dphi$, yields the solution in regimes 2--4 in the supplemental material of Ref.~\cite{Gus21}: $\dphi=\tfrac{Y_{gp}}{Y_{gg}}$ with $Y_{gp}$ given by contracting Eq.~(S18) in Ref.~\cite{Gus21} with $\pphat$.
The solution to Eq.~(\ref{eq:dphiODEfirst}) generalizes the solution in Ref.~\cite{Gus21} by introducing non-zero correction to the force due to fluid inertia.
It is compared to the result based on the full solution (\ref{eq:dphiODEsolution}) in Fig.~\ref{fig:orientation_phi_comparison}.
It agrees fairly well for flow F1, with larger deviations for the flows with larger Reynolds number.

Second, in the limit $\st\to 0$, Eq.~(\ref{eq:dphiODEfirst}) simplifies to
\begin{align}
-\sigma\tfrac{{\rm d}}{{\rm d}t}\dphi
-(\sigma+\Delta)Y_{gg}\dphi
=\sigma B_{gp}(t).
\end{align}
Since $\sv$ is large (so $Y_{gg}$ is large), this system quickly relaxes to its instantaneous fixed point
\begin{align}
\dphi^*=-\frac{\sigma}{(\sigma+\Delta)Y_{gg}} B_{gp}(t)
=\frac{\tfrac{C_\perp}{I_\perp}\Ap B_{gp}(t)\Big(1+C_f\tfrac{3a_\perp}{16}(3\Ap-\Ao)v_g^*\Big)}{C_\tau{\mathscr A}' A_\perp[v_g^*]^2\Big(1 + C_f\tfrac{3a_\perp}{8}A_\perp v_g^*\Big)}\,.
\label{eq:dphiStar}
\end{align}
This fixed point was used to derive the theory for $\langle\dphi^2\rangle$ for prolate spheroids in turbulence in Eq.~(S6) in Ref.~\cite{Bhowmick24}.
It also agrees with the fixed point $\dphi^*=\tfrac{Y_{gp}}{Y_{gg}}$ in Ref.~\cite{Gus21}, as well as the slender-body results in Refs.~\cite{Kramel,Roy2023}.
The solution (\ref{eq:dphiStar}) therefore applies to the overdamped regime 2 as defined in Ref.~\cite{Men17,Kramel,Gus21,Roy2023}.

Third, Ref.~\cite{Gus21} derived a driven oscillator equation for $\dphi$ by assuming $W_p=-u_p$.
This is a valid assumption in the limit of large $\st$, such that $v_p$ mainly fluctuates around zero.
In the limit of large $\st$ and $\sv$, the dominant contribution to Eq.~(\ref{eq:dphiODE}) becomes
\begin{align}
\tfrac{{\rm d}^3}{{\rm d}t^3}\dphi+\tfrac{1}{\st}[\tfrac{\Co}{\Io}-\sigma]\tfrac{{\rm d}^2}{{\rm d}t^2}\dphi
+\tfrac{\Co}{\Io\st}Y_{gg}\tfrac{{\rm d}}{{\rm d}t}\dphi
=-\tfrac{\Co}{\Io\st}\tfrac{Y_{gg}}{v_g^*}\tfrac{{\rm d}}{{\rm d}t}u_p(t).
\end{align}
Here we neglected the term proportional to $Y_{gg}\dphi/\st^2$ since $\dphi\sim\sv^{-2}$ is small.
Integrating this equation and setting the integration constant to zero to keep the average angle unbiased gives
\begin{align}
\tfrac{{\rm d}^2}{{\rm d}t^2}\dphi+\tfrac{1}{\st}[\tfrac{\Co}{\Io}-\sigma]\tfrac{{\rm d}}{{\rm d}t}\dphi
+\tfrac{\Co}{\Io\st}Y_{gg}\dphi
=-\tfrac{\Co}{\Io\st}\tfrac{Y_{gg}}{v_g^*}u_p(t).
\label{eq:dphiODEsecond}
\end{align}
Setting $\sigma/\st=0$, consistent with the assumption $W_p=-u_p$, yields Eq.~(S12) in Ref.~\cite{Gus21}.

We conclude by remarking that Ref.~\cite{Gus21} showed that the solution to Eq.~(\ref{eq:dphiODEfirst}) can be obtained by approximating $\dphi=\tfrac{Y_{gp}}{Y_{gg}}$ and by solving the equation for $v_p$ using this approximation.
The resulting solution is valid if the rotational dynamics is overdamped (regimes 2 and 3), but it turns out to also be accurate in the underdamped regime under a mean-field approximation (regime 4).
Here, we find a solution to the dynamics of $\dphi$, $\omega_s$ and $v_p$ by solving Eq.~(\ref{eq:dphiODE}) that does not require $\dphi=\tfrac{Y_{gp}}{Y_{gg}}$, making it valid for an even larger parameter range.

\end{document}